\documentclass[12pt,leqno]{amsart}
\usepackage{amssymb}
\usepackage{amsmath,amssymb}
\newtheorem{theorem}{Theorem}[section]

\newtheorem{proposition}{Proposition}[section]

\begin{document}
\theoremstyle{plain}
\newtheorem{MainThm}{Theorem}
\newtheorem{thm}{Theorem}[section]
\newtheorem{clry}[thm]{Corollary}
\newtheorem{prop}[thm]{Proposition}
\newtheorem{lem}[thm]{Lemma}
\newtheorem{deft}[thm]{Definition}
\newtheorem{hyp}{Assumption}
\newtheorem*{ThmLeU}{Theorem (J.~Lee, G.~Uhlmann)}

\theoremstyle{definition}
\newtheorem{rem}[thm]{Remark}
\newtheorem*{acknow}{Acknowledgments}
\numberwithin{equation}{section}
\newcommand{\eps}{{\varphi}repsilon}
\renewcommand{\d}{\partial}
\newcommand{\re}{\mathop{\rm Re} }
\newcommand{\im}{\mathop{\rm Im}}
\newcommand{\R}{\mathbf{R}}
\newcommand{\C}{\mathbf{C}}
\newcommand{\N}{\mathbf{N}}
\newcommand{\D}{C^{\infty}_0}
\renewcommand{\O}{\mathcal{O}}
\newcommand{\dbar}{\overline{\d}}
\newcommand{\supp}{\mathop{\rm supp}}
\newcommand{\abs}[1]{\lvert #1 \rvert}
\newcommand{\csubset}{\Subset}
\newcommand{\detg}{\lvert g \rvert}
\newcommand{\ppp}{\partial}
\newcommand{\dd}{\mbox{div}\thinspace}
\newcommand{\www}{\widetilde}

\title
[]
{Remark on Calder\'on's problem for the system of  elliptic equations}


\author{
O.~Yu.~Imanuvilov and \, M.~Yamamoto }

\thanks{
Department of Mathematics, Colorado State
University, 101 Weber Building, Fort Collins, CO 80523-1874, U.S.A.
e-mail: {\tt oleg@math.colostate.edu}.
Partially supported by NSF grant DMS 1312900}\,
\thanks{ Department of Mathematical Sciences, The University
of Tokyo, Komaba, Meguro, Tokyo 153, Japan e-mail:
myama@ms.u-tokyo.ac.jp}

\date{}

\maketitle


\begin{abstract}
We consider the Calder\'on problem  in the case of partial Dirichlet-to-Neumann map for the system of elliptic equations in a bounded two
dimensional domain. The main result of the manuscript is as follows:
If two systems of elliptic operators generate the same partial Dirichlet-to-Neumann map  the coefficients  can be uniquely determined up to the gauge equivalence.
\end{abstract}

\section{Introduction}\label{sec1}
Let $\Omega$ be a bounded domain in $\Bbb R^2$ with smooth boundary,
let $\tilde \Gamma$ be an open set on $\partial\Omega$ and
$\Gamma_0=Int (\partial\Omega\setminus \tilde \Gamma).$ Consider the
following boundary value problem:
\begin{equation}\label{o-1}
L(x,D)u=\Delta u+2A\partial_z u+2B\partial_{\bar z} u+Qu=0\quad
\mbox{in}\,\,\Omega,\quad  u\vert_{\Gamma_0}=0,\quad u\vert_{\tilde
\Gamma}=f.
\end{equation}
Here $u=(u_1,\dots, u_N)$ be a unknown vector function and  $A, B,
Q$ be  smooth $N\times N$ matrices. Consider the following partial
Dirichlet-to-Neumann map:
$$\Lambda_{A,B,Q}f=\partial_{\vec \nu} u,
\quad
\mbox{where}\quad
L(x,D)u=0\quad\mbox{in}\,\,\Omega,\quad u\vert_{\Gamma_0}=0,\quad
u\vert_{\tilde \Gamma}=f,
$$
where $\vec\nu$ is the outward unit normal to $\partial\Omega.$
This inverse problem is the generalization of so called Calder\'on's
problem (see \cite{C}), which itself is the mathematical realization
of {\sl Electrical Impedance Tomography} (EIT).
The goal of this paper is to extend the result obtained in \cite{ESK}  for the above problem in three-dimensional convex domain, which states that the coefficients of two systems  of elliptic equations  which principal part is the Laplace  operator and  which produce the same Dirichlet-to-Neumann map can be determined up to the gauge equivalence.

 We have

\begin{theorem}\label{vokal} Let $A_j,B_j\in C^{5+\alpha}(\bar \Omega), Q_j\in  C^{4+\alpha}(\bar \Omega)$ for
$j=1,2$  and  some $\alpha\in (0,1)$ and for the operators $L_j(x,D)$ of the form (\ref{o-1}) with coefficients  $A_j,B_j,Q_j$ and adjoint of these operators zero is not an eigenvalue. Suppose that $\Lambda_{A_1,B_1,Q_1}=\Lambda_{A_2,B_2,Q_2}.$ Then
\begin{equation}\label{op!}
A_1=A_2\quad\mbox{and}\,\, B_1=B_2\quad \mbox{on} \,\,\tilde \Gamma,
\end{equation}
and there exists an invertible matrix $\mbox{\bf Q}\in C^{5+\alpha}(\bar \Omega)$ such that
\begin{equation}
\mbox{\bf Q}\vert_{\tilde \Gamma}=I,\quad \partial_{\vec \nu}\mbox{\bf Q}\vert_{\tilde \Gamma}=0,
\end{equation}
\begin{equation}\label{A1}
A_2=2\mbox{\bf Q}^{-1}\partial_{\bar z}\mbox{\bf Q}+\mbox{\bf Q}^{-1}A_1\mbox{\bf Q}\quad\mbox{in}\,\,\Omega,
\end{equation}
\begin{equation}\label{A2}
B_2=2\mbox{\bf Q}^{-1}\partial_{z}\mbox{\bf Q}+\mbox{\bf Q}^{-1}B_1\mbox{\bf Q}\quad\mbox{in}\,\,\Omega,
\end{equation}
\begin{equation}\label{A3}
Q_2=\mbox{\bf Q}^{-1}Q_1\mbox{\bf Q}+\mbox{\bf Q}^{-1}\Delta \mbox{\bf Q}+2\mbox{\bf Q}^{-1}A_1\partial_z\mbox{\bf Q}+2\mbox{\bf Q}^{-1}B_1\partial_{\bar z}\mbox{\bf Q}\quad\mbox{in}\,\,\Omega .
\end{equation}
\end{theorem}

The paper organized as follows.
In section 3 we construct the complex geometric optics solutions for the boundary value problem (1.1). In section 4 we prove  some asymptotic   for coefficients of two operators $L_j(x,D)$ of the form (\ref{o-1}) which generate the same Dirichlet-to-Neumann map. In section 5, from the asymptotic relations obtained in the section 4, it is proved that there exists  a gauge transformation $\mbox{\bf Q}$ which preserves the Dirichlet-to-Neumann map and such that it transforms the coefficient $ A_1\rightarrow A_2.$ Then for the coefficients operators  $\mbox{\bf Q}^{-1}L_1(x,D)\mbox{\bf Q}$ and $L_2(x,D)$ we obtain some system of integral-differential equations. Finally in the section 6 we study this integral-differential equation and show that the operators $\mbox{\bf Q}^{-1}L_1(x,D)\mbox{\bf Q}$ and $L_2(x,D)$ are the same.

{\bf Notations.}
Let $i=\sqrt{-1}$ and $\overline{z}$ be the complex conjugate of
$z \in \Bbb C$. We set
$\partial_z = \frac 12(\partial_{x_1}-i\partial_{x_2})$,
$\partial_{\overline z}= \frac12(\partial_{x_1}+i\partial_{x_2})$ and
$$
\partial_{\overline z}^{-1}g=-\frac 1\pi\int_\Omega
\frac{g(\xi_1,\xi_2)}{\zeta-z}d\xi_1d\xi_2,\quad
\partial_{ z}^{-1}g=-\frac 1\pi\int_\Omega
\frac{g(\xi_1,\xi_2)}{\overline\zeta-\overline z}d\xi_1d\xi_2.
$$
For any holomorphic function $\Phi$ we set $\Phi'=\partial_z\Phi$ and $\bar\Phi'=\partial_{\bar z}\bar \Phi,$ $\Phi''= \partial^2_z\Phi$,$\bar\Phi''=\partial^2_{\bar z}\bar \Phi.$  Let $\vec \tau=(\nu_2,-\nu_1)$ be tangential vector to $\partial\Omega$. Let $W^{1,\tau}_2(\Omega)$ be  the Sobolev space $W^1_2(\Omega)$ with the norm $\Vert u\Vert_{W^{1,\tau}_2(\Omega)}=\Vert \nabla u\Vert_{L^2(\Omega)}+\vert\tau\vert\Vert u\Vert_{L^2(\Omega)}.$ Moreover by $\lim_{\eta\to\infty} \frac{\Vert f(\eta)\Vert_X}{\eta} = 0$
and $\Vert f(\eta)\Vert_X \le C\eta$ as $\eta \to \infty$ with some $C>0$,
we define $f(\eta) = o_X(\eta)$ and $f(\eta) = O_X(\eta)$ as
$\eta \to \infty$ for a normed space $X$ with norm $\Vert \cdot\Vert_X$,
respectively. $\beta=(\beta_1,\beta_2),$ $\beta_i\in\Bbb N_+,$ $\vert\beta\vert=\beta_1
+\beta_2,$ $I$ is the identity matrix.

\section{\bf  Construction of the operators $P_B$ and $T_B$.}

Let $A, B$ be an $N\times N$ matrix with elements from $C^{5+\alpha}
(\overline\Omega)$ with $\alpha\in (0,1).$ Consider the boundary value problem:
\begin{equation}\label{1-55!}
\mathcal K(x,D)(U_{0},\widetilde U_{0})=(2\partial_{\overline z}U_{0} +A
U_{0}, 2\partial_{ z}\widetilde U_{0} +B\widetilde
U_{0})=0\quad\mbox{in}\,\,\Omega,\quad U_{0}+\widetilde U_{0}=0\quad
\mbox{on}\,\,\Gamma_0.
\end{equation}

Without loss of generality we assume  that $\tilde \Gamma$  is an ark with endpoints $x_\pm.$

We have
\begin{proposition}(see \cite{IY3})\label{nikita}
Let $\epsilon$ be a positive number, $A,B\in C^{5+\alpha}(\bar \Omega)$  for some $\alpha\in (0,1),$ $\Psi\in C^\infty(\partial\Omega)$
$\vec r_{0,k},\dots ,\vec r_{2,k}\in \Bbb C^3$ be arbitrary vectors
and $x_1,\dots, x_k$ be mutually distinct arbitrary  points from the domain $\Omega.$
There exists a solution $(U_{0},\widetilde U_{0})\in C^{6+\alpha}(\overline \Omega)$ to
problem (\ref{1-55!})  such that
\begin{equation}\label{xoxo1}
\partial_z^j U_{0}(x_\ell)=\vec r_{j,\ell}\quad
\forall j\in \{0,\dots, 5\},\quad\mbox{and}\quad \forall \ell\in\{1,\dots, k\},
\end{equation}
\begin{equation}\label{xoxo1u}
\lim_{x\rightarrow x_\pm}\frac{ \vert U_0(x)\vert}{\vert x-x_\pm\vert^{98}}
=\lim_{x\rightarrow x_\pm}\frac{ \vert \widetilde U_0(x)\vert}
{\vert x-x_\pm\vert^{98}}=0
\end{equation}
and
\begin{equation}\label{gomr}
\Vert U_0-\Psi\Vert_{C^{5+\alpha}(\bar\Gamma_0)}\le \epsilon .
\end{equation}
\end{proposition}

We construct  the matrix $\mathcal C$ and  the matrix $\mathcal P$  as follows
\begin{equation}
\mathcal C=(\tilde U_0(1),\dots,\tilde U_0(N)),\,\, \mathcal P=( U_0(1),\dots, U_0(N))\in C^{6+\alpha}(\bar\Omega)
\end{equation} and for any $j\in\{1,\dots,N\}$
\begin{equation}\label{-55!U3}
\mathcal K(x,D)(U_{0}(j),\widetilde U_{0}(j))=0\quad\mbox{in}\,\,\Omega,\quad U_{0}(j)+\widetilde U_{0}(j)=0\quad
\mbox{on}\,\,\Gamma_0.
\end{equation}
By Proposition \ref{nikita} for the equation (\ref{-55!U3})
we  can construct solutions $(U_{0}{(j)},\tilde U_{0}{(j)})$ such that
$$
U_{0}{(j)}(\hat x)=\vec e_j,\quad \forall j\in\{1,\dots, N\},
$$
where $\vec e_j$ is the standard basis in $\Bbb R^N.$

By $\mathcal Z$ we denote the set of zeros of the function $q$ on $\overline
\Omega$ :  $\mathcal Z=\{z\in \overline\Omega; \, q(z)=0\}.$
Obviously $card\, \mathcal Z<\infty.$ By $\kappa$ we denote the highest order
of zeros of the function $q$ on $\overline \Omega.$

Using  Proposition 9 of \cite{IY}
we construct solutions $U_{0}^{(j)}$ to problem (\ref{-55!U}) such that
$$
U_{0}^{(j)}( x)=\vec e_j\quad \forall j\in\{1,\dots ,N\}\quad\mbox{and}\quad
\forall x\in \mathcal Z.
$$
Set $\widetilde {\mathcal P} (x)=( U_{0}^{(1)}(x),\dots ,U_{0}^{(N)}(x)),\widetilde {\mathcal C} (x)=(\widetilde U_{0}^{(1)}(x),\dots ,\widetilde U_{0}^{(N)}(x)).$ Then there exists a holomorphic function $\widetilde q$
such that $\mbox{det}\,\widetilde {\mathcal P}=\widetilde q(z)e^{-\frac 12
\partial^{-1}_{\bar z}\mbox{tr}\,\widetilde{ \mathcal P}}$ in $\Omega.$
Let $\widetilde{\mathcal Z}=\{z\in \overline\Omega; \thinspace
\widetilde q(z)=0\}$ and $\widetilde \kappa$ the highest order of zeros of
the function $\widetilde q.$

By $\widetilde U_{0}^{(j)}(x) = \vec{e_j}$ for $x \in \mathcal{Z}$, we see that
$
\widetilde{\mathcal Z}\cap\mathcal Z=\emptyset.
$
Therefore there exists a holomorphic function $r(z)$ such that
$r\vert_{\mathcal Z}=0$ and $ (1-r)
\vert_{\widetilde{\mathcal Z}}=0
$
and the orders of zeros of the function $r$ on $\mathcal Z$ and the function
$1-r$ on $\widetilde {\mathcal Z}$ are greater than or equal to the
$\max\{\kappa,\widetilde \kappa\}.$

We set
\begin{equation}\label{victory}
P_{A}f=\frac 12 \mathcal P\partial^{-1}_{\overline z} (\mathcal P^{-1}rf)+\frac 12
\widetilde {\mathcal P}\partial^{-1}_{\overline z} (\widetilde {\mathcal P}^{-1}(1-r)f).
\end{equation}
Then
$$
P_{A}^*f=-\frac 12 r (\mathcal P^{-1})^*\partial^{-1}_{\overline z} (\mathcal P^*f)
-\frac 12(1-r) (\widetilde {\mathcal P}^{-1})^*\partial^{-1}_{\overline z}
(\widetilde {\mathcal P}^*f).
$$
For any matrix $A\in C^{5+\alpha}(\overline \Omega), \alpha\in (0,1)$,  the linear operators $P_{A}, P_A^*\in {\mathcal L} (L^2(\Omega),
W^1_2(\Omega))$ solve the differential equations
\begin{equation}\label{oblom11}\nonumber
(-2\partial_{\overline z}+A^*)P^*_{A}g=g\quad \mbox{in}\,\,\Omega \quad
(2\partial_{\overline z}+A)P_{A}g=g\quad \mbox{in}\,\,\Omega.
\end{equation}

In a similar way, using  matrices $\mathcal C, \widetilde{ \mathcal C}$ we construct the  operators
\begin{equation}\label{giorgi}\nonumber
T_{B}f=\frac 12 \mathcal C\partial^{-1}_{z} (\mathcal C^{-1} \bar rf)
+ \frac 12 \widetilde{ \mathcal C}\partial^{-1}_{ z}
(\widetilde {\mathcal C}^{-1}(1-\bar r)f)
\end{equation}
and
\begin{equation}\label{giorgi1}
T_{B}^*f=-\frac 12  r(\overline z) (\mathcal C^{-1})^*\partial^{-1}_{ z}
(\mathcal C^*f)-\frac 12(1-r(\overline z)) (\widetilde {\mathcal C}^{-1})^*
\partial^{-1}_{z} (\widetilde {\mathcal C}^*f).
\end{equation}

For any matrix $B\in C^{5+\alpha}(\overline \Omega), \alpha\in (0,1)$,  the linear
operators $T_{B}$ and $T^*_{B}$ solve the differential equation
\begin{equation}\label{oblom1}\nonumber
(2\partial_z+B)T_{B}g=g\quad \mbox{in}\,\,\Omega \quad\mbox{and}\quad
(-2\partial_{ z}+B^*)T^*_{B}g=g\quad \mbox{in}\,\,\Omega.
\end{equation}

Finally we introduce two operators
\begin{equation}\label{NaNa}\nonumber
\widetilde {\mathcal R}_{\tau, B}g = e^{\tau(\overline\Phi-\Phi)}T_B
(e^{\tau(\Phi-\overline\Phi)}g) \quad\mbox{and}\quad
{\mathcal R}_{\tau, B}g = e^{\tau(\Phi-\overline\Phi)}
P_B(e^{\tau(\overline\Phi-\Phi)}g).
\end{equation}

\section{\bf Step 1: Construction of complex geometric optics solutions.}

Let $L_1(x,D)$ and $L_2(x,D)$ be the operators  of the form (\ref{o-1}) with the coefficients $A
_j,B_j, Q_j.$
In this step, we will construct two complex geometric optics solutions
$u_1$ and $v$ respectively for operators $L_1(x,D)$ and $L_2(x,D).$

%
%

As the phase function for such a solution  we consider a holomorphic
function $\Phi(z)$ such that $ \Phi(z)
=\varphi(x_1,x_2)+i\psi(x_1,x_2) $  with real-valued functions
$\varphi$ and
$\psi.$ For some $\alpha\in (0,1)$ the function $\Phi$ belongs to
$C^{6+\alpha}(\overline{\Omega}).$ Moreover
\begin{equation}\label{zzz}
\partial_{\bar z}\Phi = 0 \quad \mbox{in}
\,\,\Omega, \quad\mbox{Im}\,\Phi\vert_{\Gamma_0}=0.
\end{equation}
 Denote by $\mathcal H$ the set
of all the critical points of the function $\Phi$:
$
\mathcal H = \{z\in\overline\Omega; \thinspace
\Phi' (z)=0\}.
$
Assume that $\Phi$ has no critical points on
$\overline{\widetilde\Gamma}$, and that all critical points  are
nondegenerate:
\begin{equation}\label{mika}
\mathcal H\cap \partial\Omega=\emptyset,\quad
\Phi'' (z)\ne 0, \quad \forall z\in
\mathcal H,\quad\mbox{card}\, \mathcal H<\infty.
\end{equation}

Let $\partial\Omega=\cup_{j=1}^{\mathcal N}\gamma_j.$
The following proposition
 was proved
in \cite{IUY}.

\begin{proposition}\label{Proposition -1}
Let $\widetilde x$ be an arbitrary point in  domain $\Omega.$ There exists a
sequence of functions $\{\Phi_\epsilon\}_{\epsilon\in(0,1)}\in C^6(\bar\Omega)$
satisfying (\ref{zzz}), (\ref{mika}
) and there exists a sequence
$\{\widetilde x_\epsilon\}, \epsilon\in (0,1)$ such that
\begin{equation}\label{bobik1}
\widetilde x_\epsilon \in \mathcal H_\epsilon
= \{z\in\overline\Omega; \thinspace
\Phi_\epsilon'(z)=0 \},\quad \widetilde x_\epsilon\rightarrow \widetilde
x\,\,\mbox{ as}\,\, \epsilon\rightarrow +0.
\end{equation}
and
\begin{equation}\label{bobik2}
\mbox{Im}\,\Phi_\epsilon(\widetilde x_\epsilon)\notin \{\mbox{Im}\,
\Phi_\epsilon(x); \thinspace x\in \mathcal H_\epsilon\setminus
\{\widetilde{x_\epsilon}\}\} \,\,\mbox{and}
\,\,\mbox{Im}\,\Phi_\epsilon(\widetilde x_\epsilon) \ne 0.
\end{equation}
\end{proposition}

Let the function $\Phi$ satisfy (\ref{zzz}), (\ref{mika}) and $\widetilde x$
be some point from $\mathcal H.$
Without loss of generality, we may assume that $\widetilde \Gamma$ is an arc
with the endpoints $x_\pm.$

Denote
$
Q_1(1)=-2\partial_z A_1-B_1A_1+Q_1,\quad Q_2(1)=-2\partial_{\overline z}
B_1-A_1B_1+Q_1.
$

Let $(U_{0},\widetilde U_{0})\in C^{6+\alpha}(\overline \Omega)$
be a  solution to the boundary value problem:
\begin{equation}\label{-55!}
\mathcal K(x,D)(U_{0},\widetilde U_{0})=(2\partial_{\overline z}U_{0} +A_1
U_{0}, 2\partial_{ z}\widetilde U_{0} +B_1 \widetilde
U_{0})=0\quad\mbox{in}\,\,\Omega,\quad U_{0}+\widetilde U_{0}=0\quad
\mbox{on}\,\,\Gamma_0.
\end{equation}

The complex geometric optics solutions are constructed in  \cite{IY},  \cite{IY3}. We remind the main steps.
Let the pair $(U_0,\widetilde U_0)$ be defined in the following way
\begin{equation}
U_0=\mathcal P_1\mbox{\bf a},\quad \tilde U_0=\mathcal C_1\overline{\mbox{\bf a}},
\end{equation}
where $\mbox{\bf a}(z)=(a_1(z),\dots,a_N(z))\in C^{5+\alpha}(\bar \Omega)$ is the holomorphic vector function such that $\mbox{Im}\,\,\mbox{\bf a}\vert_{\Gamma_0}=0,$ or
\begin{equation}
U_0=\mathcal P_1\mbox{\bf a},\quad \tilde U_0=-\mathcal C_1\overline{\mbox{\bf a}},
\end{equation}
where $\mbox{\bf a}(z)=(a_1(z),\dots,a_N(z))\in C^{5+\alpha}(\bar \Omega)$ is the holomorphic vector function such that $\mbox{Re}\,\,\mbox{\bf a}\vert_{\Gamma_0}=0,$
\begin{equation}\label{Ax}
\mathcal C_1=(\tilde U_0(1),\dots,\tilde U_0(N)),\,\, \mathcal P_1=( U_0(1),\dots, U_0(N))\in C^{6+\alpha}(\bar\Omega)
\end{equation} and  for any $k\in\{1,\dots,N\}$
\begin{equation}\label{-55!U}
\mathcal K(x,D)(U_{0}(k),\widetilde U_{0}(k))=0\quad\mbox{in}\,\,\Omega,\quad U_{0}(k)+\widetilde U_{0}(k)=0\quad
\mbox{on}\,\,\Gamma_0.
\end{equation}
In order to make a choice of $\mathcal C_1,\mathcal P_1$
let us  fix  a small positive number  $\epsilon.$
By Proposition \ref{nikita} there exist solutions  $(U_0(k),\tilde U_0(k))$  to problem (\ref{-55!}) for  $k\in\{1,\dots, N\}$ such that
\begin{equation}\label{kino}
\Vert U_0(k)-\vec e_k\Vert_{C^{5
+\alpha}(\bar\Gamma_0)}\le \epsilon\quad\forall k\in\{1,\dots,N\}.
\end{equation}
This inequality and the boundary conditions in (\ref{-55!})  on $\Gamma_0$ imply
\begin{equation}\label{kinogovno}
\Vert \tilde U_0(k)-\vec e_k\Vert_{C^{5
+\alpha}(\bar\Gamma_0)}\le \epsilon\quad\forall k\in\{1,\dots,N\}.
\end{equation}

Let $e_1,e_2$ be smooth functions such that
\begin{equation}\label{short}
 e_1+e_2=1\quad
\mbox{on}\,\,\Omega,\quad
\end{equation}
and $e_1$ vanishes in a neighborhood of $\partial\Omega$ and $e_2$
vanishes in a neighborhood of the set $\mathcal H.$

For any positive $\epsilon$ denote $G_\epsilon=\{x\in \Omega; \thinspace
dist (\mbox{supp}\,e_1,x)>\epsilon\}.$  The following proposition proved in \cite{IY}:
\begin{proposition} \label{popo}
Let  $B, q\in C^{5+\alpha}(\overline\Omega)$  for some positive
$\alpha\in (0,1),$ the function $\Phi$ satisfy (\ref{zzz}), (\ref{mika})  and $\widetilde q\in W^1_p(\overline\Omega)$ for some
$p>2.$ Suppose that $q\vert_{\mathcal H}=\widetilde q\vert_{\mathcal
H}=0.$  Then the
asymptotic formulae hold true:
\begin{eqnarray}\label{50}
\widetilde{\mathcal R}_{\tau, B}(e_1(q+\frac{\widetilde q}{\tau}))\vert_{\overline
G_\epsilon}
= e^{\tau(\overline\Phi-\Phi)}\left (\frac{m_{+,\tilde x} e^{2i\tau\psi
(\widetilde x)}}{\tau^2}+o_{C^2(\overline
G_\epsilon)}(\frac{1}{\tau^2})\right)
\quad\mbox{as}\,\vert\tau\vert\rightarrow +\infty ,\\
\quad{\mathcal R}_{\tau, B} (e_1(q + \frac{\widetilde q}{\tau}))\vert_{\overline
G_\epsilon}
=e^{\tau(\Phi-\overline\Phi)}\left (\frac{m_{-,\tilde x}
e^{-2i\tau\psi(\widetilde x)}}{\tau^2} +o_{C^2(\overline
G_\epsilon)}(\frac{1}{\tau^2})\right
)\quad\mbox{as}\,\vert\tau\vert\rightarrow +\infty.
\end{eqnarray}
\end{proposition}

Denote $q_1=P_{A_1}(Q_1(1)U_{0})-M_1,\quad q_2=
T_{B_1}(Q_2(1)\widetilde
U_{0})-M_2\in C^{5+\alpha}(\bar \Omega),$
 where the functions
$M_1\in Ker (2\partial_{\overline z}+A_1)$ and $M_2\in Ker
(2\partial_z+B_1) $ are taken such that
\begin{equation}\label{kl}
q_1(\tilde x)=q_2(\tilde x)=\partial^\beta_xq_1(x)=\partial^\beta_xq_2(x)=0, \quad
\forall x\in \mathcal H\setminus\{\tilde x\}\quad \mbox{and}\,\,\forall\,\vert \beta\vert\le 5.
\end{equation}
Moreover by (\ref{xoxo1u}) we can assume that
\begin{equation}\label{bin1}
\lim_{x\rightarrow x_\pm}\frac{ \vert q_1(x)\vert}{\vert x-x_\pm\vert^{98}}
=\lim_{x\rightarrow x_\pm}\frac{ \vert q_2(x)\vert}
{\vert x-x_\pm\vert^{98}}=0.
\end{equation}

Next we introduce the functions $(U_{-1},\widetilde U_{-1})\in C^{5+\alpha}(\bar\Omega)\times C^{5+\alpha}(\bar\Omega)$ as a solutions to the following
boundary value problem:
\begin{equation}\label{zad-1}
\mathcal K(x,D)(U_{-1},\widetilde U_{-1})=0\quad\mbox{in}\,\,\Omega,\\\quad
(U_{-1}+\widetilde
U_{-1})\vert_{\Gamma_0}=\frac{q_{1}}{2{\Phi '}}
+ \frac{q_{2}}{2{\bar\Phi '}}.
\end{equation}

We set $p_1=-Q_2(1)(\frac{e_1q_1}{2{\Phi '}}-U_{-1})+L_1(x,D)
(\frac{e_2q_1}{2{\Phi '}})$, $p_2=-Q_1(1)(\frac{e_1q_2}{2{\bar\Phi '}}-\widetilde
U_{-1})+L_1(x,D) (\frac{e_2q_2}{2{\bar\Phi '}}),$
$\widetilde q_2=T_{B_1}p_2-\widetilde M_2, \widetilde q_1=P_{A_1}p_1
-\widetilde M_1\in C^{5+\alpha}(\bar\Omega)$, where  $\widetilde M_1\in Ker (2\partial_{\overline z}+A_1)$
and $\widetilde M_2\in Ker (2\partial_z+B_1)$ are taken such that
\begin{equation}\label{gandon1}
\partial^\beta_x\tilde q_1(x)=\partial^\beta_x\tilde q_2(x)=0, \quad
\forall x\in \mathcal H\quad \mbox{and}\,\,\forall\vert \beta\vert\le 5.
\end{equation}

By Proposition \ref{popo}, there exist functions $m_{\pm,\tilde x}\in
C^{2+\alpha}(\overline
G_\epsilon)$
such that
\begin{equation}\label{50l}
\widetilde{\mathcal R}_{\tau, B_1}(e_1(q_1+\frac{\widetilde
q_1}{\tau})) \vert_{\overline
G_\epsilon}=
e^{\tau(\overline\Phi-\Phi)}\left (\frac{m_{+,\tilde x} e^{2i\tau\psi
(\widetilde
x)}}{\tau^2}+o_{C^2(\overline
G_\epsilon)}(\frac{1}{\tau^2})\right)
\quad\mbox{as}\,\vert\tau\vert\rightarrow +\infty
\end{equation}
and
\begin{equation}\label{50ll}
\quad{\mathcal R}_{\tau, A_1} (e_1(q_2 + \frac{\widetilde
q_2}{\tau})) \vert_{\overline
G_\epsilon} =e^{\tau(\Phi-\overline\Phi)}
\left (\frac{m_{-,\tilde x} e^{-2i\tau\psi(\widetilde x)}}{\tau^2}
+o_{C^2(\overline
G_\epsilon)}(\frac{1}{\tau^2})\right
)\quad\mbox{as}\,\vert\tau\vert\rightarrow +\infty.
\end{equation}

For any $\tilde x\in \mathcal H$ we introduce the functions
$a_{\pm,\tilde x},b_{\pm,\tilde x}\in C^{2+\alpha}(\overline \Omega)$ as solutions to the boundary value problem
\begin{equation}\label{lobster}
\mathcal K(x,D)(a_{\pm,\tilde x},b_{\pm,\tilde x})=0\quad\mbox{in}\,\,\Omega,\quad
(a_{\pm,\tilde x}+b_{\pm,\tilde x})\vert_{\Gamma_0}=m_{\pm,\tilde x}.
\end{equation}
We choose the functions $a_{\pm,\tilde x},b_{\pm,\tilde x}$ in the form
\begin{equation}\label{begemot1}
(a_{\pm,\tilde x},b_{\pm,\tilde x})=(\mathcal P_1(x)\mbox{\bf a}_{\pm,\tilde x}(z),\mathcal C_1(x)\mbox{\bf b}_{\pm,\tilde x}(\bar z)),
\end{equation}
where $\mbox{\bf a}_{\pm,\tilde x}(z)$ is some holomorphic function  and $\mbox{\bf b}_{\pm,\tilde x}(\bar z)$ is some antiholomorphic function.
Let $(U_{-2},\widetilde U_{-2})
\in C^{5+\alpha}(\overline\Omega)\times  C^{5+\alpha}(\overline\Omega) $ be solution  to the
boundary value problem
\begin{eqnarray}\label{zad-11} \mathcal K(x,D)(U_{-2},\widetilde U_{-2})
=0\quad\mbox{in}\,\,\Omega,\quad
(U_{-2}+\widetilde U_{-2})\vert_{\Gamma_0}=\frac{\widetilde
q_1}{2{\Phi '}}+\frac{\widetilde q_2}{2\bar\Phi'}.\nonumber
\end{eqnarray}

We introduce the functions $U_{0,\tau}, \widetilde U_{0,\tau}
\in C^{2+\alpha}(\overline\Omega)$ by
\begin{equation}\label{zad1}
U_{0,\tau}=U_0+\frac{U_{-1}-e_2q_1/2\Phi'}{\tau}+\frac{1}{\tau^2}(( e^{2i\tau\psi(\widetilde
x)}a_{+,\tilde x}+e^{-2i\tau\psi(\widetilde x)}a_{-,\tilde x})+U_{-2}-\frac{\widetilde
q_1 e_2}{2{\Phi '}})
\end{equation}
and
\begin{equation}\label{zad2}
\widetilde U_{0,\tau}=\widetilde U_0+\frac{\widetilde U_{-1}-e_2
q_2/2{\bar\Phi '}}{\tau}+\frac{1}{\tau^2}((
e^{2i\tau\psi(\widetilde x)}b_{+,\tilde x}+e^{-2i\tau\psi(\widetilde
x)}b_{-,\tilde x})+\widetilde U_{-2}-\frac{\widetilde q_2 e_2}{2
\overline\Phi'}).
\end{equation}

We set ${\mathcal O}_{\epsilon}=\{x\in \Omega; \thinspace dist (x,\partial
\Omega)\le \epsilon\}.$

In \cite {IY}  it is shown that there exists
a function $u_{-1}$ in complex geometric
optics solution satisfies the estimate
\begin{equation}
\root\of{\vert\tau\vert} \Vert u_{-1} \Vert_{L^2(\Omega)} +
\frac{1}{\root\of{\vert\tau\vert} } \Vert \nabla u_{-1}
\Vert_{L^2(\Omega)}+\Vert u_{-1} \Vert_{W_2^{1,\tau}(\mathcal
O_{\epsilon})}=o(\frac{1}{\tau})\quad \mbox{as}\,\,\tau\rightarrow
+\infty
\end{equation}
and such that the function
\begin{equation}\label{zad}
u_1(x)=U_{0,\tau}e^{\tau \Phi}+\widetilde U_{0,\tau} e^{\tau \overline
\Phi}-e^{\tau\Phi}\widetilde{\mathcal R}_{\tau, B_1}(e_1(q_1+\widetilde
q_1/\tau))-e^{\tau\overline\Phi}{\mathcal R}_{\tau, A_1}(e_1(q_2+\widetilde
q_2/\tau))+e^{\tau \varphi} u_{-1}
\end{equation}
solves the boundary value problem
\begin{equation}\label{zad22}
L_1(x,D)u_1=0\quad\mbox{in}\,\,\Omega,\quad u_1\vert_{\Gamma_0}=0.
\end{equation}

Similarly, we construct the complex geometric optics solutions to
the operator $L_2(x,D)^*.$
Let $(V_0,\widetilde V_0) \in C^{6+\alpha}(\overline \Omega)\times C^{6+\alpha}(\overline \Omega)$
be a solutions to the following boundary value problem:
\begin{equation}\label{ll1}
\mathcal M(x,D)(V_{0},\widetilde V_{0})=((2\partial_{ z}-{B_2^*})
V_{0},(2\partial_{\overline
z}-{A_2^*})\widetilde V_{0})=0\quad\mbox{in}\,\,\Omega, \quad  (V_{0}+\widetilde
V_{0})\vert_{\Gamma_0}=0,
\end{equation}
such that
\begin{equation}\label{iiii}
\lim_{x\rightarrow x_\pm}\frac{ \vert V_0(x)\vert}{\vert x-x_\pm\vert^{98}}
=\lim_{x\rightarrow x_\pm}\frac{ \vert \widetilde V_0(x)\vert}
{\vert x-x_\pm\vert^{98}}=0.
\end{equation}

Such a pair $(V_0,\widetilde V_0)$ exists due to Proposition
\ref{nikita}. More specifically let

\begin{equation}
V_0=\mathcal C_2 \overline{ \mbox{\bf b}},\quad \tilde V_0=\mathcal P_2 \mbox{\bf b},
\end{equation}
where $\mbox{\bf b}(z)=(b_1(z),\dots,b_N(z))\in C^{5+\alpha}(\bar \Omega)$ is the holomorphic vector function such that $\mbox{Im}\,\,\mbox{\bf b}\vert_{\Gamma_0}=0,$ or
\begin{equation}
V_0=\mathcal C_2 \overline{ \mbox{\bf b}},\quad \tilde V_0=-\mathcal P_2 \mbox{\bf b},
\end{equation}
where $\mbox{\bf b}(z)=(b_1(z),\dots,b_N(z))\in C^{5+\alpha}(\bar \Omega)$ is the holomorphic vector function such that $\mbox{Re}\,\,\mbox{\bf b}\vert_{\Gamma_0}=0,$
\begin{equation}\label{ox}\mathcal C_2=(V_0(1),\dots , V_0(N)) ,\quad \mathcal P_2= (\tilde V_0(1),\dots ,\tilde V_0(N)),
\end{equation} and  for any $k\in\{1,\dots,N\}$
\begin{equation}\label{All1}
\mathcal M(x,D)(V_{0}(k),\widetilde V_{0}(k))=0\quad\mbox{in}\,\,\Omega, \quad  (V_{0}(k)+\widetilde
V_{0}(k))\vert_{\Gamma_0}=0.
\end{equation}
Moreover,
by Proposition \ref{nikita} there exist solutions  $(V_0(k),\tilde V_0(k))$  to problem (\ref{ll1}) for  $k\in\{1,\dots, N\}$ such that
\begin{equation}\label{kino1}
\Vert \tilde V_0(k)-\vec e_k\Vert_{C^{5
+\alpha}(\bar\Gamma_0)}\le \epsilon\quad\forall k\in\{1,\dots,N\}.
\end{equation}
This inequality and the boundary conditions in (\ref{ll1}) on $\Gamma_0$ imply
\begin{equation}\label{kinogovno}
\Vert  V_0(k)-\vec e_k\Vert_{C^{5
+\alpha}(\bar\Gamma_0)}\le \epsilon\quad\forall k\in\{1,\dots,N\}.
\end{equation}

In order to fix the choice of the operators  $P_{-
B^*_{2}},
T_{-A^*_{2}}$ we take $\mathcal C=\mathcal C_2, \mathcal P=\mathcal P_2$ and $\widetilde {\mathcal C}=\widetilde{\mathcal C_2}, \widetilde{\mathcal P}=\widetilde{\mathcal P_2}.$
We set
$
q_3=P_{- A^*_{2}}(Q_1(2)\widetilde V_{0})-M_{3},$ $q_4=T_{- B^*_{2}}(Q_2(2)V_{0})-M_{4}\in C^{5+\alpha}(\bar \Omega),$
where
$
Q_1(2)=Q_2^*-2\partial_{\bar z}B_2^*-B_2^*A_2^*,\quad Q_2(2)=Q_2^*-2\partial_zA_2^*-A_2^*B_2^*
$ and  $M_{3}\in Ker
(2\partial_{\overline z}-A_{2}^*)$ and
$M_{4}\in Ker (2\partial_{ z}-B_{2}^*)$ are chosen such that
\begin{eqnarray}\label{lada1}
q_{3}(\tilde x)=q_{4}(\tilde x)=\partial^\beta_x q_{3}( x)=\partial^\beta_x q_{4}( x)=0,\quad
 \forall x\in \mathcal H\setminus\{\tilde x\}\quad \mbox{and}\,\,\forall\vert\beta\vert\le 5;\nonumber\\\,\, \lim_{x\rightarrow x_\pm}\frac{ \vert q_j(x)\vert}
{\vert x-x_\pm\vert^{98}}=0\quad\forall j\in\{3,4\}.
\end{eqnarray}

By (\ref{mika}) the functions
$\frac{q_3}{2{\Phi '}},\frac{q_4}{2\overline\Phi'}$ belong to the space $C^{5+\alpha}(\overline\Gamma_0).$
 Therefore  we  can
introduce the functions $V_{-1}, \widetilde V_{-1}\in C^{5+\alpha}(\overline\Omega)$ as a solutions
to the following boundary value problem:
\begin{equation}
\mathcal M(x,D)(V_{-1},\widetilde V_{-1})=0\quad\mbox{in}\,\,\Omega,\quad
(V_{-1}+\widetilde
V_{-1})\vert_{\Gamma_0}=-(\frac{q_3}{2{\Phi '}}+\frac{q_4}
{2{\bar\Phi '}}).
\end{equation}

Let
$
p_3=Q_1(2)(\frac{e_1q_{3}}{2{\Phi '}}
+\widetilde V_{-1})+L_2(x,D)^*(\frac{q_{3}e_2}{2{\Phi '}}),
p_4=Q_2(2)
(\frac{e_1q_{4}}{2\overline\Phi'}+ V_{-1})
+L_2(x,D)^*(\frac{q_{4}e_2}{2\overline\Phi'})
$
and
$
\widetilde q_4=(T_{-
B^*_{2}}p_4-\widetilde M_{3}), \quad
\widetilde q_3=(P_{-A^*_{2}} p_3-\widetilde M_{4})\in C^{5+\alpha}(\overline\Omega),
$
where $\widetilde M_{3}\in Ker (2\partial_{ \bar z}-B_{2}^*),
\widetilde M_{4}\in Ker
(2\partial_{ z}-A_{2}^*),$ and $(\widetilde q_{3},\widetilde
q_{4})$ are chosen such that
 \begin{equation}\label{lada}
\partial^\beta_x\widetilde q_{3}( x)=\partial^\beta_x\widetilde q_{4}( x)=0,\quad
 \forall x\in \mathcal H\quad \mbox{and}\,\,\forall\vert\beta\vert\le 5.
\end{equation}

By Proposition \ref{popo},
there exist smooth functions $\widetilde m_{\pm,\tilde x}\in C^{2+\alpha}(\overline
G_\epsilon), \tilde x\in \mathcal H$,
independent of $\tau$  such that
\begin{equation}
\widetilde{\mathcal R}_{-\tau,- B^*_{2}}(e_1(q_{3}+\widetilde
q_{3}/\tau))\vert_{\bar G_\epsilon}=\frac{\widetilde m_{+,\tilde x}
e^{2i\tau(\psi-\psi(\widetilde
x))}}{\tau^2}+e^{2i\tau\psi}o_{C^2(\overline{ G_\epsilon})}(\frac{1}{\tau^2})\quad\mbox{as}\,\,\vert\tau\vert\rightarrow
+\infty
\end{equation}
and
\begin{equation}
{\mathcal R}_{-\tau,- A_{2}^*}(e_1(q_{4}+\widetilde
q_{4}/\tau))\vert_{\bar G_\epsilon}=\frac{\widetilde m_{-,\tilde x}
e^{-2i\tau(\psi-\psi(\widetilde
x))}}{\tau^2}+e^{-2i\tau\psi}o_{C^2(\overline{ G_\epsilon})}(\frac{1}{\tau^2})\quad\mbox{as}\,\,\vert\tau\vert\rightarrow
+\infty.
\end{equation}

Using the functions $\tilde m_{\pm,\tilde x} $ we introduce functions
$\widetilde
 a_{\pm,\tilde x}, \widetilde  b_{\pm,\tilde x} \in C^{2+\alpha}(\overline \Omega)$ which solve the boundary value problem

\begin{equation}\label{anakonda}
\mathcal M(x,D)(\widetilde a_{\pm,\tilde x},\widetilde
b_{\pm,\tilde x})=0\quad\mbox{in}\,\,\Omega,\quad (\widetilde a_{\pm,\tilde x}+\widetilde
b_{\pm,\tilde x})\vert_{\Gamma_0}=\widetilde m_{\pm,\tilde x} .
\end{equation}
We choose $\widetilde a_{\pm,\tilde x},\widetilde b_{\pm,\tilde x}$ in the form
\begin{equation}\label{begemot2}
(\widetilde a_{\pm,\tilde x},\widetilde b_{\pm,\tilde x})=(\mathcal C_2(x)\widetilde{\mbox{\bf a}}_{\pm,\tilde x}(\bar z),\mathcal P_2(x)\widetilde{\mbox{\bf b}}_{\pm,\tilde x}(z)),
\end{equation}
where $\mbox{\bf a}_{\pm,\tilde x}(\bar z)$ is some antiholomorphic function  and $\mbox{\bf b}_{\pm,\tilde x}(z)$ is some holomorphic function.
By (\ref{mika}) the functions
$\frac{\widetilde q_3}{2{\Phi '}},\frac{\widetilde q_4}{2\overline\Phi'}$ belong to the space $C^{5+\alpha}(\overline\Gamma_0).$ Therefore there exists a pair $(V_{-2},\widetilde V_{-2})\in
C^{5+\alpha}(\bar\Omega)\times C^{5+\alpha}(\bar\Omega)$ which solves  the boundary value problem
\begin{equation}
\mathcal M(x,D)(V_{-2},\widetilde V_{-2})=0\quad
\mbox{in}\,\,\Omega,
\quad
(V_{-2}+\widetilde V_{-2})\vert_{\Gamma_0}=-(\frac{\widetilde
q_3}{2{\Phi '}}+\frac{\widetilde q_4}{2
\overline\Phi'}).
\end{equation}

We introduce  functions $V_{0,\tau},
\widetilde V_{0,\tau}\in C^{2+\alpha}(\bar\Omega)$ by formulas
\begin{equation}\label{-1}
\tilde V_{0,\tau}=\tilde V_{0}+\frac{\tilde V_{-1}+e_2q_{3}/2
\Phi'}{\tau}+\frac{1}{\tau^2}( e^{2i\tau\psi(\widetilde x)}\widetilde
b_{+,\tilde x}+e^{-2i\tau\psi(\widetilde x)}\widetilde
b_{-,\tilde x}+\tilde V_{-2}+\frac{e_2\widetilde q_{3}}{2{\Phi '}})
\end{equation}
and
\begin{equation}\label{-2}
V_{0,\tau}=V_{0}+\frac{
V_{-1}+e_2q_{4}/2\overline
\Phi'}{\tau}+\frac{1}{\tau^2}( e^{2i\tau\psi(\widetilde x)}\widetilde
a_{+,\tilde x}+e^{-2i\tau\psi(\widetilde x)}\widetilde a_{-,\tilde x}+
V_{-2}+\frac{e_2\widetilde q_{4}}{2{\bar\Phi '}}).
\end{equation}

The last term $v_{-1}$ in
complex geometric optics solution satisfies the estimate
\begin{equation}\label{Amimino11}
\root\of{\vert\tau\vert} \Vert v_{-1} \Vert_{L^2(\Omega)} +
\frac{1}{\root\of{\vert\tau\vert} } \Vert \nabla v_{-1}
\Vert_{L^2(\Omega)}+\Vert v_{-1} \Vert_{W_2^{1,\tau}(\mathcal
O_{\epsilon})}=o(\frac{1}{\tau})\quad \mbox{as}\,\, \tau\rightarrow+\infty
\end{equation}
and such that the function
\begin{equation}\label{-3}
v=V_{0,\tau}e^{-\tau \bar\Phi}+\widetilde V_{0,\tau}e^{-\tau
\Phi}-e^{-\tau\Phi}\widetilde{\mathcal R}_{-\tau,-
B^*_{2}}(e_1(q_{3}+\frac{\widetilde
q_{3}}{\tau}))
-e^{-\tau\overline\Phi}{\mathcal R}_{-\tau,
-A_{2}^*}(e_1(q_{4}+\frac{\widetilde q_{4}}{\tau}))
+v_{-1}e^{-\tau \varphi}
\end{equation}
solves the boundary value problem
\begin{equation}\label{-4}
L_2(x,D)^*v=0\quad\mbox{in}\,\,\Omega, \quad  v\vert_{\Gamma_0}=0.
\end{equation}

We close this section with one technical proposition similar to one  proved in \cite{IUY2}:

\begin{proposition}\label{balda} Suppose that the  functions $\mathcal C_i,\mathcal P_i\in C^{6+\alpha}(\bar\Omega)$ for all $ i,j\in \{1,2\}$ some $\alpha\in (0,1)$  given by (\ref{Ax})-(\ref{kino}), (\ref{ox})-(\ref{kino1}) satisfy
\begin{equation}\label{1C}
\int_{\partial\Omega}\{(\nu_1+i\nu_2)\Phi'(\mathcal P_1 {\mbox{\bf a}},\mathcal P_2 {\mbox{\bf b}})+(\nu_1-i\nu_2)\bar\Phi'(\mathcal C_1\bar{\mbox{\bf a}},\mathcal C_2\bar{\mbox{\bf b}})\}d\sigma=0,
\end{equation}
 for all holomorphic vector  functions ${\mbox{\bf a}}, {\mbox{\bf b}}$ such that $\mbox{Im}\, {\mbox{\bf a}}\vert_{\Gamma_0}=\mbox{Im}\, {\mbox{\bf b}}\vert_{\Gamma_0}=0.$  Then there exist a holomorphic function
$\Theta\in W_2^\frac 12(\Omega)$ and an antiholomorphic function
$\widetilde \Theta\in W_2^\frac 12(\Omega)$ such that
\begin{equation}\label{volt}
\widetilde\Theta\vert_{\widetilde \Gamma}=\mathcal C_2^* \mathcal C_1,
\quad \Theta\vert_{\widetilde \Gamma}=\mathcal P_2^*\mathcal P_1
\end{equation}
and
\begin{equation}\label{ix}
\Theta=\tilde \Theta \quad\mbox{on}\quad\Gamma_0.
\end{equation}
\end{proposition}

{\bf Proof.} First we show that  for  all holomorphic vector  functions ${\mbox{\bf a}}, {\mbox{\bf b}}$ such that $\mbox{Im}\, {\mbox{\bf a}}\vert_{\Gamma_0}=\mbox{Im}\, {\mbox{\bf b}}\vert_{\Gamma_0}=0 $ there exists a holomorphic function $ \widetilde\Psi$ and antiholomorphic function  $\Psi$ such that
$$
 \bar \Phi'(\mathcal C_1\bar {\mbox{\bf a}},\mathcal C_2\bar {\mbox{\bf b}})-\Psi=\Phi'(\mathcal P_1{\mbox{\bf a}},\mathcal P_2{\mbox{\bf b}})-{\widetilde\Psi}=0\quad\mbox{on}\,\,\tilde\Gamma\,\,\mbox{and}\,\,((\nu_1-i\nu_2)\Psi+(\nu_1+i\nu_2)\widetilde \Psi)
\vert_{\Gamma_0}=0.
 $$
 Also we observe that the equality (\ref{1C}) implies \begin{equation}\label{1CC}
\mathcal I=\int_{\partial\Omega}\{(\nu_1+i\nu_2)\Phi'(\mathcal P_1 {\mbox{\bf a}},\mathcal P_2 {\mbox{\bf b}})+(\nu_1-i\nu_2)\bar\Phi'(\mathcal C_1(-\bar{\mbox{\bf a}}),\mathcal C_2\bar{\mbox{\bf b}})\}d\sigma=0,
\end{equation}
for  all holomorphic vector  functions ${\mbox{\bf a}}, {\mbox{\bf b}}$ such that $\mbox{Re}\, {\mbox{\bf a}}\vert_{\Gamma_0}=\mbox{Im}\, {\mbox{\bf b}}\vert_{\Gamma_0}=0. $
Indeed,
$$\mathcal I=\frac 1 i\int_{\partial\Omega}\{(\nu_1+i\nu_2)\Phi'(\mathcal P_1 {i\mbox{\bf a}},\mathcal P_2 {\mbox{\bf b}})+(\nu_1-i\nu_2)\bar\Phi'(\mathcal C_1(-i\bar{\mbox{\bf a}}),\mathcal C_2\bar{\mbox{\bf b}})\}d\sigma=
$$
$$
\frac 1i\int_{\partial\Omega}\{(\nu_1+i\nu_2)\Phi'(\mathcal P_1 {i\mbox{\bf a}},\mathcal P_2 {\mbox{\bf b}})+(\nu_1-i\nu_2)\bar\Phi'(\mathcal C_1(\overline{i\mbox{\bf a}}),\mathcal C_2\overline{\mbox{\bf b}})\}d\sigma=0.
$$
Here, in order to get the last equality we used (\ref{1C}).
Consider the extremal problem:
\begin{equation}\label{AX}
J(\Psi,\widetilde \Psi)=\Vert \bar \Phi'(\mathcal C_1\bar {\mbox{\bf a}},\mathcal C_2\bar {\mbox{\bf b}})-\Psi\Vert^2
_{L^2(\widetilde \Gamma)}
+\Vert  \Phi'(\mathcal P_1{\mbox{\bf a}},\mathcal P_2{\mbox{\bf b}})-{\widetilde\Psi}\Vert^2_{L^2(\widetilde \Gamma)}\rightarrow \inf,
\end{equation}
\begin{equation}\label{voron}
\frac{\partial\Psi}{\partial z}=0\quad\mbox{in}\,\Omega,
\quad \frac{\partial\widetilde \Psi}{\partial\overline z}=0\quad\mbox{in}\,\Omega,
\quad ((\nu_1-i\nu_2)\Psi+(\nu_1+i\nu_2)\widetilde \Psi)
\vert_{\Gamma_0}=0.
\end{equation}

Denote the unique solution to this extremal problem (\ref{AX}),
(\ref{voron}) by $(\widehat \Psi,\widehat{ \widetilde \Psi})$. Applying the
Fermat theorem, we obtain
\begin{equation}\label{ipoa}
\mbox{Re} (\Phi'(\mathcal P_1 {\mbox{\bf a}},\mathcal P_2 {\mbox{\bf b}})-\widehat{\tilde \Psi},\delta)_{L^2(\widetilde\Gamma)}
+\mbox{Re}
(\bar\Phi'(\mathcal C_1\bar{\mbox{\bf a}},\mathcal C_2\bar{\mbox{\bf b}})-\widehat { \Psi},
\widetilde \delta)_{L^2(\widetilde\Gamma)}=0\quad
\end{equation}
for any $\delta,\widetilde\delta$  from $ W_2^\frac 12(\Omega)$ such that
\begin{equation}\label{voron1}
\,\,\frac{\partial\delta}{\partial \bar z} =0 \quad\mbox{in}\,\Omega,
\quad \frac{\partial\widetilde \delta}{\partial z}= 0\quad\mbox{in}
\,\Omega,\quad (\nu_1+i\nu_2)\delta\vert_{\Gamma_0}=-(\nu_1-i\nu_2)
\widetilde \delta\vert_{\Gamma_0}
\end{equation}
and there exist two functions $P,\widetilde P\in W_2^\frac 12(\Omega)$ such that
\begin{equation}\label{020}
\frac{\partial P}{\partial \overline z}=0\quad\mbox{in}\,\,\Omega,\quad
\frac{\partial\widetilde P}{\partial z}=0\quad\mbox{in}\,\,\Omega,
\end{equation}
\begin{equation}\label{021}
(\nu_1+i\nu_2) P=\overline{\Phi'(\mathcal P_1 {\mbox{\bf a}},\mathcal P_2 {\mbox{\bf b}})-\widehat {\tilde\Psi}}\quad\mbox{on}
\,\widetilde \Gamma,\quad (\nu_1-i\nu_2) \widetilde P
=\overline{\bar\Phi'(\mathcal C_1\bar{\mbox{\bf a}},\mathcal C_2\bar{\mbox{\bf b}})
-\widehat {\Psi}}\quad\mbox{on}\,\widetilde \Gamma
\end{equation}
and
\begin{equation}\label{ona}
(P-\widetilde P)\vert_{\Gamma_0}=0.
\end{equation}
Denote $\Psi_0(z)=\frac{1}{2i}(P(z)
-\overline{\widetilde P(\overline z)})$ and
$\Phi_0(z)=\frac 12(P(z)+\overline {\widetilde P(\overline z)}).$
Equality (\ref{ona}) yields
\begin{equation}\label{chech}
\mbox{Im}\,\Psi_0\vert_{\Gamma_0}=\mbox{Im}\,\Phi_0\vert_{\Gamma_0}=0.
\end{equation}
Hence
\begin{equation}\label{vodka}
P=(\Phi_0+i\Psi_0), \quad\overline {\widetilde P}=(\Phi_0-i\Psi_0).
\end{equation}
From (\ref{ipoa}), taking $\delta=\widehat \Psi$ and $\widetilde \delta
=\widehat{\widetilde \Psi}$, we have
\begin{equation}\label{Pona}
\mbox{Re} \int_{\tilde \Gamma}
((\bar\Phi'(\mathcal C_1\bar{\mbox{\bf a}},\mathcal C_2\bar{\mbox{\bf b}})-\widehat { \Psi},\overline{\widehat { \Psi}})+(\Phi'(\mathcal P_1 {\mbox{\bf a}},\mathcal P_2 {\mbox{\bf b}})-\widehat{ \tilde\Psi},\overline{\widehat {\tilde\Psi}}))
d\sigma
=0.
\end{equation}
By (\ref{020}), (\ref{021}) and (\ref{vodka}), we have
\begin{eqnarray}
H_1=\mbox{Re}\int_{\tilde\Gamma} (\overline{\Phi'(\mathcal P_1 {\mbox{\bf a}},\mathcal P_2 {\mbox{\bf b}})-\widehat{\tilde \Psi}},{\Phi'(\mathcal P_1 {\mbox{\bf a}},\mathcal P_2 {\mbox{\bf b}})})
+ (\overline{\bar\Phi'(\mathcal C_1 \bar {\mbox{\bf a}},\mathcal C_2 \bar{\mbox{\bf b}})
-\widehat { \Psi}},{\bar\Phi'( \mathcal C_1 \bar{\mbox{\bf a}},\mathcal C_2 \bar{\mbox{\bf b}})})
d\sigma\nonumber\\
=\mbox{Re}\int_{\tilde\Gamma}((\nu_1+i\nu_2) P,{\Phi'(\mathcal P_1 {\mbox{\bf a}},\mathcal P_2 {\mbox{\bf b}})})
+ ((\nu_1-i\nu_2){\widetilde P},
{\bar\Phi'(\mathcal C_1\bar{\mbox{\bf a}},\mathcal C_2\bar{\mbox{\bf b}})})
d\sigma
                                   =\nonumber\\
\mbox{Re}\int_{\tilde\Gamma}2 ((\nu_1+i\nu_2){(\Phi_0+i\Psi_0)}
{\Phi'(\mathcal P_1 {\mbox{\bf a}},\mathcal P_2 {\mbox{\bf b}})})+2
((\nu_1-i\nu_2){(\bar\Phi_0-\overline{i\Psi_0})}
{\bar\Phi'(\mathcal C_1\bar{\mbox{\bf a}},\mathcal C_2\bar{\mbox{\bf b}})})
d\sigma.
                                          \nonumber
\end{eqnarray}
By (\ref{1C}) and (\ref{chech}) we have
\begin{equation}\label{nona}
\int_{\tilde\Gamma}2\mbox{Re} ((\nu_1-i\nu_2)\overline{\Phi_0
\Phi'(\mathcal P_1 {\mbox{\bf a}},\mathcal P_2 {\mbox{\bf b}})})+2\mbox{Re}
((\nu_1+i\nu_2) \overline{\bar\Phi_0\bar\Phi'(\mathcal C_1\bar{\mbox{\bf a}},\mathcal C_2\bar{\mbox{\bf b}})})
d\sigma=0.
\end{equation}

By (\ref{1CC}) and (\ref{chech}) we obtain
\begin{equation}\label{nona1A}
\int_{\tilde\Gamma}2\mbox{Re} ((\nu_1+i\nu_2){(i\Psi_0)
\Phi'(\mathcal P_1 {\mbox{\bf a}},\mathcal P_2 {\mbox{\bf b}})})
+2\mbox{Re} ((\nu_1-i\nu_2) {}{\bar\Phi'(\mathcal C_1\overline{(-i \Psi_0)\mbox{\bf a}},\mathcal C_2\bar{\mbox{\bf b}})})d\sigma=0.
\end{equation}

Then by  (\ref{nona}) and (\ref{nona1A})  we see that $H_1=0.$
Taking into account (\ref{Pona}), we obtain that $J(\widehat \Psi,
\widehat{\widetilde\Psi})=0.$
Hence
\begin{equation}\label{voron3}
(\mathcal P_1 {\mbox{\bf a}},\mathcal P_2 {\mbox{\bf b}})(x)=(\tilde\Psi/\Phi')(z)=\tilde\Xi(z),\quad (\mathcal C_1 \bar{\mbox{\bf a}},\mathcal C_2\bar {\mbox{\bf b}})(x)=(\Psi/\bar\Phi')(\bar z)=\Xi (\bar z)\quad \mbox{on}\,\,\tilde \Gamma.
\end{equation}
 In general the function $\Phi$ may have a finite number of zeros
in $\Omega.$ At these zeros $\Xi,\widetilde \Xi$
may have poles.  On the other hand observe
that $\Xi,\widetilde \Xi$ are independent of
a particular  choice of the function $\Phi.$ Making small perturbations of these functions, we can shift
the position of the zeros of the function $\Phi'.$ Hence there are no poles for $\Xi,\widetilde \Xi.$
By (\ref{voron}) $((\nu_1-i\nu_2)\Psi+(\nu_1+i\nu_2)\tilde \Psi)\vert_{\Gamma_0}=0.$ Moreover, by the direct computations, $(\nu_1+i\nu_2)\Phi'+(\nu_1-i\nu_2)\bar\Phi')\vert_{\Gamma_0}=0$ . Therefore
\begin{equation}\label{voron2}
\tilde\Xi(z)= \Xi(\bar z)\quad\mbox{on}\,\,\Gamma_0.
\end{equation}
 Consider $N$ holomorphic  vector functions ${\mbox{\bf b}}_j=({\mbox{\bf b}}_{1,j},\dots , {\mbox{\bf b}}_{1,N})$ such that $Im\,{\mbox{\bf b}}_j\vert_{\Gamma_0}=0$ and determinant of the square matrix constructed from these vector functions  not equal to zero at least at one point of domain $\Omega.$ Then equality (\ref{voron3}) can be written as
$$
(\mathcal P_2^*\mathcal P_1 {\mbox{\bf a}},{\mbox{\bf b}}_j)={\tilde\Xi}_j (z)\quad\mbox{and}\quad (\mathcal C_2^*\mathcal C_1 \bar{\mbox{\bf a}},\bar {\mbox{\bf b}}_j)=\Xi_j (\bar z)\quad \mbox{on}\,\,\tilde \Gamma.
$$
Then
$$
\mathcal P_2^*\mathcal P_1 {\mbox{\bf a}}=\mbox{\bf B}^{-1}\vec{\tilde \Xi}\quad \mbox{and}\quad\mathcal C_2^*\mathcal C_1 \bar{\mbox{\bf a}}=\bar {\mbox{\bf B}}^{-1}\vec {\Xi}\quad \mbox{on}\,\,\tilde \Gamma.
$$
Here $B$ is the matrix such that  the row  number $j$ equal ${\mbox{\bf b}}_j^t$ and $\vec {\tilde\Xi}(z)=({\tilde\Xi}_1 (z),\dots, {\tilde\Xi}_N (z)),\vec{ \Xi}=(\Xi_1 (\bar z),\dots,\Xi_N (\bar z)).$
Consider $N$ holomorphic  vector functions ${\mbox{\bf a}}_i$ such that $Im\,{\mbox{\bf a}}_i\vert_{\Gamma_0}=0.$
Then
$$
\mathcal P_2^*\mathcal P_1 {\mbox{\bf a}}_i=\mbox{\bf B}^{-1}\vec {\tilde\Xi_i}\quad \mbox{and}\quad\mathcal C_2^*\mathcal C_1 \bar{\mbox{\bf a}}_i=\bar{\mbox{\bf  B}}^{-1}\vec{ \Xi}_i\quad \mbox{on}\,\,\tilde \Gamma.
$$
From this equality we have
$$
\mathcal P_2^*\mathcal P_1 =\mbox{\bf B}^{-1}\Pi \mbox{\bf A}^{-1}\quad \mbox{and}\quad\mathcal C_2^*\mathcal C_1=\bar{\mbox{\bf B}}^{-1}\tilde \Pi \bar {\mbox{\bf A}}^{-1}\quad \mbox{on}\,\,\tilde \Gamma.
$$
Here $\mbox{\bf A}, \Pi, \tilde \Pi $ are matrix such that  the row  number $i$ equal ${\mbox{\bf a}}_i, \vec \Xi_i$ and $\vec {\tilde \Xi}_i.$ We set
$$\Theta=\mbox{\bf B}^{-1}\Pi \mbox{\bf A}^{-1}\quad \mbox{and }\quad\tilde\Theta=\bar {\mbox{\bf B}}^{-1}\tilde \Pi \bar{\mbox{\bf  A}}^{-1}.$$ These formulae defines the functions $\Theta, \tilde \Theta$ correctly except the point where determinants of matrix $\mbox{\bf A}$ and $\mbox{\bf B}$ are equal to zero. On the other hand it is obvious  that functions $\Theta,\tilde\Theta$ are independent of the choice  of matrices $\mbox{\bf A}, \mbox{\bf B}.$ So if we assume that there exist  a point of singularity of, say, the function $\Theta$ by Proposition \ref{nikita} we can make a choice matrices $\mbox{\bf A},\mbox{\bf B}$ such that determinants of these  matrices do not equal to zero at this point and arrive to the contradiction.  The equality (\ref{ix}) follows from (\ref{voron2}) and the  fact that $Im\,\mbox{\bf B}\vert_{\Gamma_0}= Im\,\mbox{\bf A}\vert_{\Gamma_0}=0.$ Indeed on $\Gamma_0$
$$\mathcal P_2^*\mathcal P_1 =\mbox{\bf B}^{-1}\Pi \mbox{\bf A}^{-1}=\bar{\mbox{\bf B}}^{-1}\Pi \bar{\mbox{\bf  A}}^{-1}=\bar{\mbox{\bf  B}}^{-1}\tilde \Pi \bar{\mbox{\bf A}}^{-1}=\mathcal C_2^*\mathcal C_1.
$$
Proof of the  proposition is complete. $\blacksquare$

 Let $u_1$ be the complex geometric optics solution given by (\ref{zad}) constructed for the operator $L_1(x,D).$ Since the Dirichlet-to-Neumann maps for the operators $L_1(x,D)$ and $L_2(x,D)$ are equal
there exists a  function   $ u_2$ be a solution to the following boundary value
problem:
$$
{ L}_{2}(x,D) u_2=0\quad \mbox{in}\,\,\Omega,\quad
 ( u_1-u_2)\vert_{\partial\Omega}=0, \quad
\partial_{\vec\nu}( u_1-u_2)=0\quad \mbox{on}\,\,\tilde \Gamma.
$$

Setting $ u= u_1-u_2$ we have
\begin{equation}
{L}_2(x,{D}) u+2\mathcal A\partial_z u_1
+2{\mathcal B}\partial_{\overline z} u_1
+{\mathcal Q} u_1=0 \quad \mbox{in}~ \Omega,    \label{Mn}
\end{equation}
where $\mathcal A=A_1-A_2,\mathcal B=B_1-B_2$ and $\mathcal Q=Q_1-Q_2$ and
\begin{equation}\label{mn11}
u \vert_{\partial\Omega} =0, \quad \partial_{\vec \nu} u
\vert_{\widetilde \Gamma} =0.
\end{equation}
Let $v$ be a function given by  (\ref{-3}).  Taking the scalar
product of (\ref{Mn}) with $ v$  in $L^2(\Omega)$ and using
(\ref{-4}) and (\ref{mn11}), we obtain
\begin{equation}\label{ippolit}
0= \int_{\Omega}(2{\mathcal A}\partial_{z} u_1 +2{\mathcal B}\partial_{\overline z} u_1 +{\mathcal Q}u_1,
v) dx.
\end{equation}
Denote
\begin{equation}\label{gnomik1}
V=V_{0,\tau}e^{-\tau \bar\Phi}+\widetilde V_{0,\tau}e^{-\tau
\Phi}-e^{-\tau\Phi}\widetilde{\mathcal R}_{-\tau,-
B^*_{2}}(e_1(q_{3}+\frac{\widetilde
q_{3}}{\tau}))
-e^{-\tau\overline\Phi}{\mathcal R}_{-\tau,
-A_{2}^*}(e_1(q_{4}+\frac{\widetilde q_{4}}{\tau}))
\end{equation} and

\begin{equation}\label{gnomik}
U=U_{0,\tau}e^{\tau \Phi}+\widetilde U_{0,\tau} e^{\tau \overline
\Phi}-e^{\tau\Phi}\widetilde{\mathcal R}_{\tau,  B_1}(e_1(q_1+\widetilde
q_1/\tau))-e^{\tau\overline\Phi}{\mathcal R}_{\tau, A_1}(e_1(q_2+\widetilde
q_2/\tau)).
\end{equation}
We have
\begin{proposition}\label{kauk}
Let  $u_1$  is given by  (\ref{zad})
and $v$ is given by  (\ref{-3}). Then the following asymptotic holds true
$$
\int_{\Omega}(2{\mathcal A}\partial_{z} u_1 +2{\mathcal B}\partial_{\overline z} u_1 +{\mathcal Q} u_1,
v) dx=\int_{\Omega}(2\mathcal A\partial_zU+2{\mathcal B}\partial_{\overline z}U +{\mathcal Q} U,
V) dx+o(\frac 1\tau)\quad \mbox{as}\quad\tau\rightarrow +\infty,
$$
where functions $U,V$ are determined by (\ref{gnomik}) and (\ref{gnomik1}).

\end{proposition}
Proof of Proposition \ref{kauk} is exactly the same as the proof of Proposition 5.1 from  \cite{IY3}.
\bigskip
\section{\bf Step 2: Asymptotic}
\bigskip
We introduce the following functionals
$$
\frak F_{\tau,\tilde x}u=\frac{\pi}{2\vert \mbox{det}\,\psi''(\tilde x)\vert^\frac 12}\left(\frac{u(\widetilde x)}{\tau}-\frac{\partial_{zz}^2u(\widetilde x)}{2\Phi''(\tilde x)\tau^2}+\frac{\partial^2_{\overline z\overline z}u(\widetilde x)}{2\bar\Phi''(\tilde x)\tau^2}+\frac{\partial_{z}u(\widetilde x)\Phi'''(\tilde x)}{2(\Phi''(\tilde x))^2\tau^2}-\frac{\partial_{\bar z}u(\widetilde x)\bar\Phi'''(\tilde x)}{2(\bar\Phi''(\tilde x))^2\tau^2}\right)
$$
 and
$$
\frak I_\tau u=\int_{\partial\Omega} u\frac{(\nu_1-i\nu_2)}
{2\tau {\Phi '}}e^{\tau(\Phi-\overline \Phi)}d\sigma
-\int_{\partial\Omega} \frac{(\nu_1-i\nu_2)}{{\Phi '}}\partial_z
\left (\frac{u}{2\tau^2 {\Phi '}}\right )e^{\tau(\Phi-\overline \Phi)} d\sigma.
$$

Using these  notations and the  fact that $\Phi$ is the harmonic function we rewrite the classical result  of theorem 7.7.5 of \cite{Her} as
\begin{proposition}\label{osel}
Let $\Phi(z)$ satisfies (\ref{zzz}), (\ref{mika}) and $u\in C^{5+\alpha}(\bar\Omega),
\alpha\in (0,1)$ be some function.
Then the following asymptotic formula is true:
\begin{equation}\label{murzik9}
\int_{\Omega}ue^{\tau(\Phi-\overline \Phi)}dx=\sum_{\tilde y\in\mathcal H}e^{2i\tau\psi(\tilde y)}\frak F_{\tau,\tilde y}u+\frak J_{\tau}u
+ o\left(\frac{1}{\tau}\right)\quad \mbox{as}\,\,\tau\rightarrow +\infty.
\end{equation}
\end{proposition}

Denote
$$
\mbox{\bf H}(x,\partial_z,\partial_{\overline z})=2\mathcal A\partial_z+2\mathcal B\partial_{\bar z}+\mathcal Q\quad\mbox{and}\quad\mathcal J_\tau=\int_\Omega (\mbox{\bf H}(x,\partial_z,\partial_{\overline z})U,V)dx.
$$  where
 $U$ and $V$ are given by (\ref{gnomik}) and (\ref{gnomik1}) respectively.
We have
\begin{proposition}\label{lodka} The following  asymptotic holds true
\begin{eqnarray}\label{zaika}
0=\sum_{k=-1}^1\tau^k J_k+\frac 1\tau((J_++I_{+,\Phi}+K_+)(\tilde x)e^{2\tau i\psi(\tilde x)}+(J_-+ I_{-,\Phi}+K_-)(\tilde x)e^{-2\tau i\psi(\tilde x)})\nonumber\\+\int_{\tilde\Gamma}((\nu_1-i\nu_2)(\mathcal A U_0e^{\tau\Phi},V_0e^{-\tau\bar\Phi})+(\nu_1+i\nu_2)(\mathcal B \tilde U_0e^{\tau\bar\Phi},\tilde V_0e^{-\tau\Phi}))d\sigma\nonumber\\
+o(\frac 1\tau)\quad\mbox{as}\,\,\tau\rightarrow +\infty,
\end{eqnarray}
where
\begin{equation}
J_1=\int_{\partial\Omega}((\nu_1-i\nu_2)\bar\Phi'(\tilde U_0,V_0)+(\nu_1+i\nu_2)\Phi'( U_0,\tilde V_0))d\sigma,
\end{equation}
\begin{equation}
J_+(\tilde x)=\frac{\pi}{2\vert \mbox{det}\,\psi''(\tilde x)\vert^\frac 12}(-(2\partial_z\mathcal AU_0, V_0)-(\mathcal AU_0, A_2^*V_0) -(\mathcal B A_1 U_0, V_0)
+(\mathcal Q U_0,V_0))(\tilde x),
\end{equation}
\begin{equation}
J_-(\tilde x)=\frac{\pi}{2\vert \mbox{det}\,\psi''(\tilde x)\vert^\frac 12}(-(\mathcal A B_1\tilde U_0, \tilde V_0)-(2\partial_{\bar z}\mathcal B\tilde U_0, \tilde V_0)-(\mathcal B\tilde U_0, B_2^*\tilde V_0)+ (\mathcal Q\tilde U_0,\tilde V_0))(\tilde x),
\end{equation}
\begin{eqnarray}\label{lin1}
I_{\pm,\Phi}(\tilde x)=-\int_{\partial\Omega}\left\{ (\nu_1-i\nu_2)((2 b_{\pm,\tilde x}\bar\Phi', V_0)+
(2\bar\Phi'\tilde U_0, \tilde a_{\pm,\tilde x}))\right.\nonumber\\\left.+(\nu_1+i\nu_2)
((2 a_{\pm,\tilde x}\Phi', \tilde V_0)
+(2\Phi'U_0, \tilde b_{\pm,\tilde x}))\right\}d\sigma,
\end{eqnarray}
\begin{eqnarray}\label{zubilo}
K_+=
\tau\frak F_{\tau,\tilde x}(q_1 , T^*_{B_1}(B_1^*\mathcal A^* V_0)-\mathcal A^*V_0+2T_{B_1}^*(\partial_z \mathcal B^* V_0)+T_{B_1}^*(\mathcal B^*(A^*_2V_0-2\tau\bar\Phi'V_0)))
\nonumber\\
-2\tau\frak F_{\tau,\tilde x}(P_{-A_2^*}^*(\mathcal A (\partial_zU_0+\tau\Phi' U_0)+\mathcal B \partial_{\bar z} U_{0,\tau}), q_4),
\end{eqnarray}
\begin{eqnarray}\label{zubilo}
K_-=
\tau\frak F_{-\tau,\tilde x} ( q_2,P_{A_1}^*(2\partial_z(\mathcal A^*\tilde V_0) -\tau\Phi'2\mathcal A^*\tilde V_0)-\mathcal B^*\tilde V_0+P^*_{A_1}(A_1^*\mathcal B^*\tilde V_0))\nonumber\\-2\tau\frak F_{-\tau,\tilde x}( q_3,T_{-B_2^*}^*(\mathcal A\partial_z \tilde U_{0}+\mathcal B(\partial_{\bar z}\tilde U_{0}+\tau\bar\Phi'\tilde U_{0}))).
\end{eqnarray}
\end{proposition}{\bf Proof.}
By Proposition \ref{kauk}
$$
\mathcal J_\tau=\int_\Omega (\mbox{\bf H}(x,\partial_z,\partial_{\overline z})U,V)dx=o(\frac 1\tau)\quad \mbox{as}\,\,\tau\rightarrow +\infty
$$
Denote
\begin{equation}\label{pravda3}
U_1=-\widetilde{\mathcal R}_{\tau, B_1}(e_1(q_1+\widetilde
q_1/\tau)),\quad \tilde U_1= -{\mathcal R}_{\tau,
A_1}(e_1(q_2+\widetilde q_2/\tau)),
\end{equation}
\begin{equation}\label{pravda4}\tilde V_1= -\widetilde{\mathcal R}_{-\tau,-
B^*_{2}}(e_1(q_{3}+{\widetilde
q_{3}}/{\tau})),
\quad
 V_1=
-{\mathcal R}_{-\tau,
-A_{2}^*}(e_1(q_{4}+{\widetilde q_{4}}/{\tau})).
\end{equation}
Integrating by parts and using Proposition \ref{osel}, we obtain

\begin{eqnarray}
\mathcal M_1=\int_\Omega (2\mathcal A\partial_z (U_{0,\tau}e^{\tau\Phi}) +2\mathcal B\partial_{\bar z} (U_{0,\tau}e^{\tau\Phi}), V_{0,\tau}e^{-\tau\bar\Phi})dx=\nonumber\\
\int_\Omega ((-2\partial_z\mathcal AU_{0,\tau}e^{\tau\Phi}, V_{0,\tau}e^{-\tau\bar\Phi})-(2\mathcal A U_{0,\tau}e^{\tau\Phi}, \partial_zV_{0,\tau}e^{-\tau\bar\Phi}) +(2\mathcal B\partial_{\bar z} U_{0,\tau}e^{\tau\Phi}, V_{0,\tau}e^{-\tau\bar\Phi}))dx\nonumber\\
+\int_{\partial\Omega}(\nu_1-i\nu_2)(\mathcal A U_{0,\tau}e^{\tau\Phi},V_{0,\tau}e^{-\tau\bar\Phi})d\sigma\nonumber=\\
e^{2i\tau\psi(\tilde x)}\frak F_{\tau,\tilde x}(-(2\partial_z\mathcal AU_0, V_0)-(2\mathcal AU_0, \partial_zV_0) +(2\mathcal B\partial_{\bar z} U_0, V_0))\nonumber\\+
\frak I_\tau(-(2\partial_z\mathcal AU_{0,\tau}, V_{0,\tau})-(2\mathcal AU_{0,\tau}, \partial_zV_{0,\tau}) +(2\mathcal B\partial_{\bar z} U_{0,\tau}, V_{0,\tau}))\nonumber\\
+\int_{\tilde \Gamma}(\nu_1-i\nu_2)(\mathcal A U_0,V_0)e^{\tau(\Phi-\bar\Phi)}d\sigma+\kappa_{0,0}+\frac{\kappa_{0,-1}}{\tau}+o(\frac 1\tau),
\end{eqnarray}
where $\kappa_{0,j}$ are  some constants independent of $\tau.$

Integrating by parts we obtain that there exist constants  $\kappa_{1,j}$  independent of $\tau$ such that
\begin{eqnarray}\label{sor1}
\int_\Omega (2\mathcal A\partial_z (\tilde U_{0,\tau}e^{\tau\bar\Phi}) +2\mathcal B\partial_{\bar z} (\tilde U_{0,\tau}e^{\tau\bar \Phi}), V_{0,\tau}e^{-\tau\bar\Phi})dx=\nonumber\\
(2\mathcal A\partial_z\tilde U_{0,\tau},V_{0,\tau})_{L^2(\Omega)}+(2\mathcal B(\partial_z\tilde U_{0,\tau}+\tau\bar \Phi'\tilde U_{0,\tau}),V_{0,\tau})_{L^2(\Omega)}=\nonumber\\
\tau\kappa_{1,1}+\kappa_{1,0}+\frac{\kappa_{1,-1}}{\tau}+\frac 1\tau(e^{2i\tau\psi(\tilde x)}(2\mathcal B b_{+,\tilde x}\bar\Phi', V_0)_{L^2(\Omega)}+e^{-2i\tau\psi(\tilde x)}(2\mathcal B b_{-,\tilde x}\bar\Phi', V_0)_{L^2(\Omega)})\nonumber\\
+\frac 1\tau(e^{2i\tau\psi(\tilde x)}(2\mathcal B\bar\Phi' \tilde U_0, \tilde a_{+,\tilde x})_{L^2(\Omega)}+e^{-2i\tau\psi(\tilde x)}(2\mathcal B\bar\Phi'\tilde U_0, \tilde a_{-,\tilde x})_{L^2(\Omega)})+o(\frac 1\tau).
\end{eqnarray}
Since by (\ref{-55!}), (\ref{lobster}), (\ref{ll1}), (\ref{anakonda}) for  the functions $\tilde a_{\pm,\tilde x}, b_{\pm,\tilde x}$ we have
$$(2\mathcal B\bar\Phi' \tilde U_0, \tilde a_{\pm,\tilde x})=-4\partial_z(\bar\Phi' \tilde U_0, \tilde a_{\pm,\tilde x}),\quad\mbox{and}\,\,(2\mathcal B b_{\pm,\tilde x}\bar\Phi', V_0)=-4\partial_z(b_{\pm,\tilde x}\bar\Phi', V_0)\quad\mbox{in}\,\,\Omega
$$ from (\ref{sor1}) we have
\begin{eqnarray}\label{sor2}
\mathcal M_2=\int_\Omega (2\mathcal A\partial_z (\tilde U_{0,\tau}e^{\tau\bar\Phi}) +2\mathcal B\partial_{\bar z} (\tilde U_{0,\tau}e^{\tau\bar \Phi}), V_{0,\tau}e^{-\tau\bar\Phi})dx=\nonumber\\
\tau\kappa_{1,1}+\kappa_{1,0}+\frac{\kappa_{1,-1}}{\tau}+\int_{\partial\Omega}\frac {(\nu_1-i\nu_2)}{\tau}(e^{2i\tau\psi(\tilde x)}(2\mathcal B b_{+,\tilde x}\bar\Phi', V_0)+e^{-2i\tau\psi(\tilde x)}(2\mathcal B b_{-,\tilde x}\bar\Phi', V_0))d\sigma\nonumber\\
+\int_{\partial\Omega}\frac {(\nu_1-i\nu_2)}{\tau}(e^{2i\tau\psi(\tilde x)}(2\bar\Phi'\tilde U_0, \tilde a_{+,\tilde x})+e^{-2i\tau\psi(\tilde x)}(2\bar\Phi'\tilde U_0, \tilde a_{-,\tilde x}))d\sigma+o(\frac 1\tau).
\end{eqnarray}
Integrating by parts we obtain that there exist  constants $\kappa_{2,j}$  independent of $\tau$ such that
\begin{eqnarray}\label{sor3}
\int_\Omega (2\mathcal A\partial_z (U_{0,\tau}e^{\tau\Phi}) +2\mathcal B\partial_{\bar z} (U_{0,\tau}e^{\tau\Phi}),\tilde V_{0,\tau}e^{-\tau\Phi})dx=\nonumber\\
(2\mathcal A(\partial_z U_{0,\tau}+\tau\Phi' U_{0,\tau}) +2\mathcal B\partial_{\bar z} U_{0,\tau},\tilde V_{0,\tau})_{L^2(\Omega)}=\nonumber\\
\tau\kappa_{2,1}+\kappa_{1,0}+\frac{\kappa_{2,-1}}{\tau}+
\frac 1\tau(e^{2i\tau\psi(\tilde x)}(2\mathcal A a_{+,\tilde x}\Phi', \tilde V_0)_{L^2(\Omega)}+e^{-2i\tau\psi(\tilde x)}(2\mathcal A a_{-,\tilde x}\Phi',\tilde  V_0)_{L^2(\Omega)})\nonumber\\
+\frac 1\tau(e^{2i\tau\psi(\tilde x)}(2\mathcal A \Phi'\tilde U_0, \tilde b_{+,\tilde x})_{L^2(\Omega)}+e^{-2i\tau\psi(\tilde x)}(2\mathcal A \Phi'\tilde U_0,\tilde  b_{-,\tilde x})_{L^2(\Omega)})+o(\frac 1\tau).
\end{eqnarray}
Since by (\ref{-55!}), (\ref{lobster}), (\ref{ll1}), (\ref{anakonda}) for  the functions $ a_{\pm,\tilde x}, \tilde b_{\pm,\tilde x}$ we have
$$
(2\mathcal A a_{\pm,\tilde x}\Phi', \tilde V_0)=-4\partial_{\bar z}( a_{\pm,\tilde x}\Phi', \tilde V_0)\quad\mbox{and}\quad (2\mathcal A \Phi'\tilde U_0, \tilde b_{\pm,\tilde x})=-4\partial_{\bar z}( \Phi'\tilde U_0, \tilde b_{\pm,\tilde x})\quad\mbox{in}\,\,\Omega
$$ we obtain from (\ref{sor3})

\begin{eqnarray}\label{sor33}
\mathcal M_3=\int_\Omega (2\mathcal A\partial_z (U_{0,\tau}e^{\tau\Phi}) +2\mathcal B\partial_{\bar z} (U_{0,\tau}e^{\tau\Phi}),\tilde V_{0,\tau}e^{-\tau\Phi})dx=\nonumber\\
\tau\kappa_{2,1}+\kappa_{1,0}+\frac{\kappa_{2,-1}}{\tau}+\int_{\partial\Omega}(\nu_1+i\nu_2)
\frac 1\tau(e^{2i\tau\psi(\tilde x)}(2 a_{+,\tilde x}\Phi', \tilde V_0)+e^{-2i\tau\psi(\tilde x)}(2 a_{-,\tilde x}\Phi',\tilde  V_0))d\sigma\nonumber\\
+\int_{\partial\Omega}(\nu_1+i\nu_2)\frac 1\tau(e^{2i\tau\psi(\tilde x)}(2\Phi'\tilde U_0, \tilde b_{+,\tilde x})+e^{-2i\tau\psi(\tilde x)}(2\Phi'\tilde U_0,\tilde  b_{-,\tilde x})) d\sigma+o(\frac 1\tau).
\end{eqnarray}
Integrating by parts, using  (\ref{-55!}) and Proposition \ref{osel},
we obtain  that there exists some constants $\kappa_{3,j}$ independent
of $\tau$ such that

\begin{eqnarray}
\mathcal M_4=\int_\Omega (2\mathcal A\partial_z (\tilde U_{0,\tau}e^{\tau\bar\Phi}) +2\mathcal B\partial_{\bar z} (\tilde U_{0,\tau}e^{\tau\bar\Phi}), \tilde V_{0,\tau}e^{-\tau\Phi})dx=\nonumber\\
\int_\Omega ((2\mathcal A\partial_z \tilde U_{0,\tau}e^{\tau\bar\Phi}, \tilde V_{0,\tau}e^{-\tau\Phi})-(2\partial_{\bar z}\mathcal B\tilde U_{0,\tau}e^{\tau\bar\Phi},\tilde V_{0,\tau}e^{-\tau\Phi}) -(2\mathcal B\tilde U_{0,\tau}e^{\tau\bar\Phi}, \partial_{\bar z}\tilde V_{0,\tau}e^{-\tau\Phi}))dx\nonumber\\
+\int_{\partial\Omega}(\nu_1+i\nu_2)(\mathcal B \tilde U_{0,\tau}e^{\tau\bar\Phi},\tilde V_{0,\tau}e^{-\tau\Phi})d\sigma\nonumber=\\
e^{-2i\tau\psi(\tilde x)}\frak F_{-\tau,\tilde x}( (2\mathcal A\partial_z \tilde U_0, \tilde V_0)-(2\partial_{\bar z}\mathcal B\tilde U_0, \tilde V_0) -(2\mathcal B\tilde  U_0, \partial_{\bar z}\tilde V_0))\nonumber\\+
\frak I_{-\tau}( (2\mathcal A\partial_z \tilde U_{0,\tau}, \tilde V_{0,\tau})-(2\partial_{\bar z}\mathcal B\tilde U_{0,\tau},\tilde  V_{0,\tau}) -(2\mathcal B \tilde U_{0,\tau}, \partial_{\bar z}\tilde V_{0,\tau}))\nonumber\\
+\int_{\tilde\Gamma}(\nu_1+i\nu_2)(\mathcal B \tilde U_0e^{\tau\bar\Phi},\tilde V_0e^{-\tau\Phi})d\sigma+\kappa_{3,1}+\frac{\kappa_{3,-1}}{\tau}+o(\frac 1\tau).
\end{eqnarray}

Integrating by parts and using Proposition \ref{osel} we obtain

\begin{eqnarray}
\mathcal M_5=\int_\Omega (2\mathcal A\partial_z (U_1e^{\tau\Phi}) +2\mathcal B\partial_{\bar z} (U_1e^{\tau\Phi}), V_{0,\tau}e^{-\tau\bar\Phi})dx=\\
\int_\Omega (\mathcal A (-B_1U_1-e_1q_1)e^{\tau\Phi} -2\partial_{\bar z}\mathcal B (U_1e^{\tau\Phi}), V_{0,\tau}e^{-\tau\bar\Phi})dx+\nonumber\\
\int_{\partial\Omega}(\nu_1+i\nu_2)(\mathcal BU_1,V_{0,\tau})e^{\tau(\Phi-\bar\Phi)}d\sigma-(2\mathcal B U_1,\partial_{\bar z}(V_{0,\tau}e^{\tau(\Phi-\bar\Phi)}))_{L^2(\Omega)}+o(\frac 1\tau)=\nonumber\\
\int_\Omega (\mathcal A (B_1T_{B_1}(e^{\tau(\Phi-\bar\Phi)}e_1q_1)-e_1q_1)e^{\tau(\Phi-\bar\Phi)}, V_{0,\tau}) +2\partial_z\mathcal B (T_{B_1}(e^{\tau(\Phi-\bar\Phi)}e_1q_1)), V_{0,\tau})dx+\nonumber\\
(\mathcal B T_{B_1}(e^{\tau(\Phi-\bar\Phi)}e_1q_1), \partial_{\bar z} V_{0,\tau}-2\tau\bar\Phi' V_{0,\tau})_{L^2(\Omega)}
+\int_{\partial\Omega}(\nu_1+i\nu_2)(\mathcal BU_1,V_{0,\tau})e^{\tau(\Phi-\bar\Phi)}d\sigma+o(\frac 1\tau)
=\nonumber\\
e^{2i\tau\psi(\tilde x
)}\frak F_{\tau,\tilde x}(q_1 , T^*_{B_1}(B_1^*\mathcal A^* V_0)-\mathcal A^*V_0+2T_{B_1}^*(\partial_z \mathcal B^* V_0)+T_{B_1}^*(\mathcal B^*(A^*_2V_0-2\tau\bar\Phi'V_0)))\nonumber\\
+\int_{\partial\Omega}(\nu_1+i\nu_2)(\mathcal BU_1,V_{0,\tau})e^{\tau(\Phi-\bar\Phi)}d\sigma+o(\frac 1\tau)\quad\mbox{as}\quad\tau\rightarrow +\infty.\nonumber
\end{eqnarray}

After integration by parts we have
\begin{eqnarray}
\mathcal M_6=\int_\Omega (2\mathcal A\partial_z (U_1e^{\tau\Phi}) +2\mathcal B\partial_{\bar z} (U_1e^{\tau\Phi}), \tilde V_{0,\tau}e^{-\tau\Phi})dx=\nonumber\\
\int_\Omega (\mathcal A (-B_1U_1-e_1q_1) -2\partial_{\bar z}\mathcal B U_1, \tilde V_{0,\tau})dx+o(\frac 1\tau)+\nonumber\\
(2\mathcal B U_1,\partial_{\bar z}\tilde V_{0,\tau})_{L^2(\Omega)}+\int_{\partial\Omega}(\nu_1+i\nu_2)(\mathcal B U_1,\tilde V_{0,\tau})d\sigma .\nonumber
\end{eqnarray}
Using (\ref{pravda3}), (\ref{50l}), (\ref{50ll}) and Proposition 8 of \cite{IY} we obtain that
\begin{equation}
\mathcal M_6=-\int_\Omega (\mathcal A q_1,\tilde V_{0,\tau})dx+o(\frac 1{\tau})\quad\mbox{as}\quad\tau\rightarrow +\infty.
\end{equation}
Integrating by parts and using Proposition \ref{osel} we have

\begin{eqnarray}\mathcal M_7=
\int_\Omega (2\mathcal A\partial_z (U_{0,\tau}e^{\tau\Phi}) +2\mathcal B\partial_{\bar z} (U_{0,\tau}e^{\tau\Phi}), V_1e^{-\tau\bar\Phi})dx=\\
2\int_\Omega (\mathcal A (\partial_zU_{0,\tau}+\tau\Phi' U_{0,\tau})e^{\tau\Phi} +\mathcal B \partial_{\bar z} U_{0,\tau}e^{\tau\Phi}, V_1e^{-\tau\bar\Phi})dx=\nonumber\\
-2\int_\Omega (P_{-A_2^*}^*(\mathcal A (\partial_zU_0+\tau\Phi' U_0)+\mathcal B \partial_{\bar z} U_{0,\tau}), e_1q_4e^{\tau(\Phi-\bar\Phi)})dx+o(\frac 1\tau)=\nonumber\\
 -2e^{2i\tau\psi(\tilde x)}\frak F_{\tau,\tilde x}(P_{-A_2^*}^*(\mathcal A (\partial_zU_0+\tau\Phi' U_0)+\mathcal B \partial_{\bar z} U_{0}), q_4)\nonumber+o(\frac 1\tau)\\
  +o(\frac 1\tau)\quad\mbox{as}\quad\tau\rightarrow +\infty.\nonumber
\end{eqnarray}

Integrating by parts and using  Proposition  8 of \cite{IY} we have
\begin{eqnarray}
\mathcal M_8=\int_\Omega (2\mathcal A\partial_z (U_{0,\tau}e^{\tau\Phi}) +2\mathcal B\partial_{\bar z} (U_{0,\tau}e^{\tau\Phi}), \tilde V_1e^{-\tau\Phi})dx=\nonumber\\
\int_\Omega( (-2\partial_z\mathcal A U_0 +\mathcal B \partial_{\bar z} U_0, \tilde V_1)-(\mathcal A U_{0,\tau},-B_2^*\tilde V_1-e_1q_3))dx+o(\frac 1\tau)\nonumber\\
+\int_{\partial\Omega}(\nu_1-i\nu_2)(\mathcal AU_0,\tilde V_1)d\sigma=-\int_\Omega(\mathcal A U_{0,\tau}, q_3)dx+o(\frac{1}{\tau})\quad\mbox{as}\quad\tau\rightarrow +\infty
\end{eqnarray}

and

\begin{eqnarray}
\mathcal M_9=\int_\Omega (2\mathcal A\partial_z (\tilde U_1e^{\tau\bar\Phi}) +2\mathcal B\partial_{\bar z} (\tilde U_1e^{\tau\bar\Phi}),  V_{0,\tau}e^{-\tau\bar\Phi})dx=\nonumber\\
\int_\Omega [(\tilde U_1, -\partial_z(2\mathcal A^* V_{0,\tau}))+(\mathcal B(-A_1\tilde U_1-e_1q_2), V_{0,\tau})]dx+o(\frac 1\tau)\nonumber\\
+\int_{\partial\Omega}(\nu_1-i\nu_2)(\mathcal A\tilde U_1, V_0)d\sigma=-\int_\Omega (\mathcal B q_2,V_{0,\tau})dx+o(\frac{1}{\tau})\quad\mbox{as}\quad\tau\rightarrow +\infty.
\end{eqnarray}

Integrating by parts and using Proposition \ref{osel} we obtain

\begin{eqnarray}
\mathcal M_{10}=\int_\Omega (2\mathcal A\partial_z (\tilde U_1e^{\tau\bar\Phi}) +2\mathcal B\partial_{\bar z} (\tilde U_1e^{\tau\bar\Phi}), \tilde  V_{0,\tau}e^{-\tau\Phi})dx=\\
\int_\Omega ((\tilde U_1, -\partial_z(2\mathcal A^* \tilde V_{0,\tau})+\tau\Phi'2\mathcal A^*\tilde V_{0,\tau})+(\mathcal B(-A_1\tilde U_1- e_1q_2),\tilde V_{0,\tau})e^{\tau(\bar\Phi-\Phi)})dx+\nonumber\\
+\int_{\partial\Omega}(\nu_1-i\nu_2)(\mathcal A\tilde U_1,\tilde V_{0,\tau})e^{\tau(\bar\Phi-\Phi)}d\sigma+o(\frac 1\tau)=\nonumber\\
\int_\Omega( e_1q_2,P_{A_1}^*(2\partial_z(\mathcal A^*\tilde V_{0,\tau}) -2\tau\Phi'\mathcal A^*\tilde V_0)-\mathcal B^*\tilde V_0+P^*_{A_1}(A_1^*\mathcal B^*\tilde V_0)))e^{\tau(\bar\Phi-\Phi)}dx\nonumber\\
+\int_{\partial\Omega}(\nu_1-i\nu_2)(\mathcal A\tilde U_1,\tilde V_0)e^{\tau(\bar\Phi-\Phi)}d\sigma+o(\frac{1}{\tau})=\nonumber\\
e^{-2i\tau\psi(\tilde x)}
+o(\frac{1}{\tau})\quad\mbox{as}\quad\tau\rightarrow +\infty.\nonumber
\end{eqnarray}

By (\ref{kl}) and Proposition \ref{osel} we obtain
\begin{eqnarray}
\mathcal M_{11}=\int_\Omega (2\mathcal A\partial_z (\tilde U_{0,\tau}e^{\tau\bar\Phi}) +2\mathcal B\partial_{\bar z} (\tilde U_{0,\tau}e^{\tau\bar\Phi}), \tilde  V_1e^{-\tau\Phi})dx=\\
\int_\Omega (2\mathcal A\partial_z \tilde U_{0,\tau}+2\mathcal B(\partial_{\bar z}\tilde U_{0,\tau}+\tau\bar\Phi'\tilde U_{0,\tau}),\tilde V_1)e^{\tau(\bar\Phi-\Phi)}dx=\nonumber\\
-\int_\Omega(e_1 q_3,T_{-B_2^*}^*(2\mathcal A\partial_z \tilde U_{0,\tau}+2\mathcal B(\partial_{\bar z}\tilde U_{0,\tau}+\tau\bar\Phi'\tilde U_{0,\tau}))e^{\tau(\bar\Phi-\Phi)}dx+o(\frac 1\tau)=\nonumber\\
-e^{-2i\tau\psi(\tilde x)}\frak F_{-\tau,\tilde x}( q_3,T_{-B_2^*}^*(2\mathcal A\partial_z \tilde U_{0}+2\mathcal B(\partial_{\bar z}\tilde U_{0}+\tau\bar\Phi'\tilde U_{0})))\nonumber\\
+o(\frac 1\tau)\quad\mbox{as}\quad\tau\rightarrow +\infty.\nonumber
\end{eqnarray}

By Proposition  \ref{osel}, there exist constants $\kappa_{4,j}$ independent of $\tau$ such that

\begin{eqnarray}
\mathcal M_{12}=\int_\Omega (\mathcal Q( U_{0,\tau}e^{\tau\bar \Phi}+ \tilde  U_{0,\tau}e^{\tau\Phi}),  V_{0,\tau}e^{-\tau\bar \Phi}+ \tilde  V_{0,\tau}e^{-\tau\Phi})dx=\\ \kappa_{4,0}+\kappa_{4,-1}/\tau+
\frac{ \pi}{ 2\tau}((\mathcal QU_0,V_0)(\tilde x)e^{2i\tau\psi(\tilde x)}+(\mathcal Q\tilde U_0,\tilde V_0)(\tilde x)e^{-2i\tau\psi(\tilde x)})+o(\frac{1}{\tau})\quad\mbox{as}\quad\tau\rightarrow +\infty.\nonumber
\end{eqnarray}
Since $\mathcal J_\tau=\sum_{k=1}^{12}\mathcal M_k$ the proof of
the Proposition \ref{lodka} is complete.
$\blacksquare$

We  have

\begin{proposition}\label{begemot} Let all conditions of the proposition \ref{osel} holds true and
\begin{equation}\label{vasilek}
A_1-A_2=B_1-B_2=0\quad\mbox{on}\,\,\tilde \Gamma.
\end{equation} For any matrices $\mathcal C_j,\mathcal P_j$ satisfying (\ref{Ax})-(\ref{kino}), (\ref{ox})-(\ref{kino1})  with sufficiently small $\epsilon$ there exists a holomorphic matrix $\Theta\in C^{5+\alpha}(\bar\Omega)$  such that the  matrix $\mbox{\bf Q}=\mathcal P_1\Theta^{-1}{\mathcal P_2^*}$  verifies
\begin{equation}\label{PLP}
2\partial_{\bar z}\mbox{\bf Q}+A_1\mbox{\bf Q}-\mbox{\bf Q}A_2=0\quad\mbox{in}\,\,\Omega\setminus \mathcal X, \quad \mbox{\bf Q}\vert_{\tilde\Gamma}=I,\quad \partial_{\vec\nu}\mbox{\bf Q}\vert_{\tilde\Gamma}=0,
\end{equation}
where $\mathcal X=\{x \in\bar\Omega\vert\mbox{det}\, \Theta=0\}$ and
\begin{equation}
\mbox{\bf Q}\in C^{6+\alpha}(\Omega\setminus\mathcal X),\quad \mbox{det}\, \mbox{\bf Q}\ne 0 \quad\mbox{in}\,\,\bar \Omega\setminus \mathcal X .
\end{equation}
%
\end{proposition}

{\bf Proof.}
From Proposition  \ref{lodka} for any  function $\Phi$ which satisfies (\ref{zzz}), (\ref{mika})  we have
\begin{equation}\label{liberation}
\int_{\partial\Omega}((\nu_1+i\nu_2)\Phi'(\tilde U_0,V_0)+(\nu_1-i\nu_2)\bar\Phi'(U_0,\tilde V_0))d\sigma=0.
\end{equation}

Then  if $\mbox{\bf a}(z)=(a_1(z),\dots,a_N(z)),\mbox{\bf  b}(z)=(b_1(z),\dots, b_N(z))$ are the holomorphic functions such that  $\mbox{Im}\, \mbox{\bf a}\vert_{\Gamma_0}=\mbox{Im}\, \mbox{\bf b}\vert_{\Gamma_0}=0$ the pairs
$(\mathcal P_1 \mbox{\bf a},\mathcal C_1\overline{\mbox{\bf a}})$ and $(\mathcal P_2 \mbox{\bf b},\mathcal C_2\overline{\mbox{\bf b}})$  solve the problems  (\ref{-55!}) and (\ref{ll1}) respectively. Therefore we can rewrite (\ref{liberation}) as
 \begin{equation}\label{liberation1}
\int_{\partial\Omega}\{(\nu_1+i\nu_2)\Phi'(\mathcal P_1 \mbox{\bf a},\mathcal P_2 \mbox{\bf b})+(\nu_1-i\nu_2)\bar\Phi'(\mathcal C_1\bar {\mbox{\bf a}},\mathcal C_2\bar {\mbox{\bf b}})\}d\sigma=0.
\end{equation}
Thanks to (\ref{liberation1}) all assumptions of the Proposition \ref{balda} holds true. By Proposition \ref{balda} there exist holomorphic  matrix $\Theta(z)$ and antiholomorphic matrix $\tilde\Theta(\bar z)$ with domain $\bar \Omega$ such that
\begin{equation}\label{govnomet}
\Theta={\mathcal P_2^*}\mathcal P_1\quad \mbox{on}\,\,\tilde \Gamma \quad\mbox{and} \quad\tilde \Theta={\mathcal C_2^*}\mathcal C_1\quad \mbox{on}\,\,\tilde \Gamma\quad \mbox{and} \quad \Theta,\tilde \Theta\in L^2(\Omega)
\end{equation}
and
\begin{equation}\label{govnomet!}
\Theta-\tilde \Theta=0\quad\mbox{on}\,\,\Gamma_0.
\end{equation}

From (\ref{kino}) and (\ref{kinogovno}) and the classical regularity theory of systems of elliptic equations (see e.g \cite{Wendland})  we obtain that $\Theta, \widetilde\Theta\in C^{6+\alpha}(\bar\Omega).$
Without loss of generality we may assume that
\begin{equation}\label{lox1}
\mbox{det}\,{\mathcal P_2^*}\ne 0\quad\mbox{and}\quad\mbox{det}\,\mathcal P_1\ne 0\quad \mbox{on}\,\,\tilde \Gamma.
\end{equation}
Moreover by (\ref{kino}), (\ref{kino1})
$$
\mbox{det}\,{\mathcal P_2^*}\ne 0\quad\mbox{and}\quad\mbox{det}\,\mathcal P_1\ne 0\quad \mbox{on}\,\, \overline\Gamma_0.
$$
 Observe that by (\ref{govnomet})
\begin{equation}\label{loshara}
I=\mathcal P_1\Theta^{-1}{\mathcal P_2^*}\quad \mbox{on}\,\,\tilde \Gamma.
\end{equation}

Since by the construction of the matrices $\mathcal P_j$
$$
2\partial_{\bar z}\mathcal P_1+A_1\mathcal P_1=0\quad \mbox{in}\,\,\Omega\quad \mbox{and}\quad 2\partial_{\bar z}{\mathcal P_2^*}-{\mathcal P_2^*}A_2=0\quad \mbox{in}\,\,\Omega
$$
and matrix $\Theta$ is holomorphic we have
$$
2\partial_{\bar z}(\mathcal P_1\Theta^{-1})+A_1(\mathcal P_1\Theta^{-1})=0\quad \mbox{in}\,\,\Omega \setminus\mathcal X.
$$
 We compute
\begin{equation}\label{gadenish}
2\partial_{\bar z}(\mathcal P_1\Theta^{-1}{\mathcal P_2^*})+A_1(\mathcal P_1\Theta^{-1}{\mathcal P_2^*})-(\mathcal P_1\Theta^{-1}{\mathcal P_2^*})A_2=0\quad\mbox{in}\,\,\Omega \setminus \mathcal X.
\end{equation}
 Thus the first equation in (\ref{PLP}) holds true. By (\ref{loshara}) the second equation  in (\ref{PLP}).

By  (\ref{vasilek}), (\ref{loshara}) on $\tilde \Gamma$ we have
\begin{equation}\label{shiroko}
-2\partial_{\bar z}{\mbox{\bf Q}}=A_1\mathcal P_1\Theta^{-1}{\mathcal P_2^*}-\mathcal P_1\Theta^{-1}{\mathcal P_2^*}A_2=A_1I- IA_2=A_1-A_2=0.
\end{equation} In order to prove the third equation in (\ref{PLP}) we observe that there exists a matrix $T(x)$ with real-valued entries, $\mbox{det}\,T(x)\ne 0,$  such that $\nabla=T(x)(\partial_{\vec\nu},\partial_{\vec \tau}).$ Therefore $\partial_{\bar z}=\frac 12((T_{11}+iT_{21})\partial_{\vec \nu}+(T_{12}+iT_{22})\partial_{\vec \tau}).$
By (\ref{shiroko}) on $\tilde \Gamma$ the following equation holds
$$
\partial_{\bar z}\mbox{\bf Q}=\frac 12((T_{11}+iT_{21})\partial_{\vec \nu}\mbox{\bf Q}+(T_{12}+iT_{22})\partial_{\vec \tau}\mbox{\bf Q})=\frac 12((T_{11}+iT_{21})\partial_{\vec \nu}\mbox{\bf Q}+(T_{12}+iT_{22})\partial_{\vec \tau}I)=
$$
$$\frac 12(T_{11}+iT_{21})\partial_{\vec \nu}\mbox{\bf Q}=0.
$$ The fact that determinant of the matrix $T$ is not equal zero implies that $(T_{11}+iT_{21})\ne 0.$ So from the above equation we have $\partial_{\vec \nu}\mbox{\bf Q}=0.$

 If $\mbox{det}\, \mbox{\bf Q}(x_0)= 0$ then $\mbox{det}\,\mathcal P_1(x_0)\mbox{det}\, \mathcal P_2(x_0)=0.$ Let  matrices $\widehat{ \mathcal P_j}$ be constructed as $\mathcal P_j$ but with the different choice  of the pairs  $(U_0(k),\tilde U_0(k)), (V_0(k),\tilde V_0(k))$ which are  solutions to problem  (\ref{-55!}) and problem (\ref{ll1}) respectively and satisfy (\ref{kino}), (\ref{kinogovno}). In such a way  we obtain another matrices  $\mathcal P_j,\Theta,\mbox{\bf Q}$ which satisfies to (\ref{PLP}) with possibly different set $\mathcal X.$ We denote such a matrix $\mathcal P_j,\Theta, \mbox{\bf Q}$ as  $\hat{\mathcal P}_j,\hat\Theta,\widehat{\mbox{\bf Q}}.$ By uniqueness of the Cauchy problem  for the $\partial_z$ operator
$$
 \mbox{\bf Q}=\widehat { \mbox{\bf Q}}\quad \mbox{on}\,\, \Omega\setminus \mathcal X\cup \widehat{\mathcal X}\quad\mbox{where}\,\,\widehat{ \mathcal X}=\{x \in\bar\Omega\vert\mbox{det} \widehat{\Theta}=0\}.
 $$
 So, $\widehat { \mbox{\bf Q}}(x_0)=0$.  On the other hand one can choose the matrices $\widehat{ \mathcal P_j}$  such that $\mbox{det}\,\widehat{ \mathcal P_j}(x_0)\ne 0.$ Therefore we arrived to the contradiction.
 Proof of the proposition is complete.
$\blacksquare$

Our next goal is to show that the matrix $\mbox{\bf Q}$ is regular on $\bar\Omega.$

Now we prove that if operators $L_j(x,D)$ generate the same Dirichlet-to-Neumann map then the operators $L_j(x,D)^*$ generate the same Dirichlet-to-Neumann map.

\begin{proposition} Let $A_j,B_j,Q_j\in C^{5+\alpha}(\bar \Omega)$ for
$j=1,2.$
If
$
\Lambda_{A_1,B_1,Q_1}=\Lambda_{A_2,B_2,Q_2}
$ then  $
\Lambda_{-A^*_1,-B^*_1,R_1}=\Lambda_{-A^*_2,-B^*_2,R_2},
$ where $R_j=-\partial_z A_j^*-\partial_{\bar z}B_j^*+Q_j^*$ for  $ j\in \{1,2\}.$
\end{proposition}

{\bf Proof.}
Let function $v_j$ solves the boundary value problem
$$
L_j(x,D)^*v_j=0\quad\mbox{in}\,\,\Omega,\quad v_j\vert_{\Gamma_0}=0,\quad v_j\vert_{\tilde \Gamma}=g
$$and $\tilde u_j$ be solution to the problem
$$
L_j(x,D)\tilde u_j=0\quad\mbox{in}\,\,\Omega,\quad \tilde u_j\vert_{\Gamma_0}=0,\quad \tilde u_j\vert_{\tilde \Gamma}=f.
$$
By our assumption and Fredholm's theorem solution for both problems exists for any $f,g\in C_0^\infty(\tilde\Gamma).$
By the Green's formula
$$
(L_j(x,D)^*v_j, \tilde u_j)_{L^2(\Omega)}-(v_j,L_j(x,D)\tilde u_j)_{L^2(\Omega)}=(\partial_{\vec \nu}v_j,\tilde u_j)_{L^2(\tilde \Gamma)}-(v_j,\partial_{\vec \nu}\tilde u_j)_{L^2(\tilde \Gamma)}+
$$
$$
(A_j(\nu_1-i\nu_2)g,f)_{L^2(\tilde \Gamma)}+(B_j(\nu_1+i\nu_2)g,f)_{L^2(\tilde \Gamma)}.
$$ Subtracting the above formulae for different $j$, using (\ref{vasilek}) and taking into account that $
\Lambda_{A_1,B_1,Q_1}=\Lambda_{A_2,B_2,Q_2}
$
we have
$$
(\partial_{\vec \nu}v_1-\partial_{\vec \nu}v_1,f)_{L^2(\tilde \Gamma)}=0.
$$
Since the function $f$ can be chosen an arbitrary from $C_0^\infty(\tilde\Gamma)$ the proof of the proposition is complete.
 $\blacksquare$

By Proposition \ref{nikita} there exist solutions  $(\mbox{U}_0(k),\tilde {\mbox U}_0(k))$  to problem
\begin{equation}\label{-155!}
(2\partial_{\overline z}{\mbox U}_{0}(k) -A^*_1
{\mbox  U}_{0}(k), 2\partial_{ z}\widetilde {\mbox U}_{0}(k) -B^*_1 \widetilde
{\mbox U}_{0}(k))=0\quad\mbox{in}\,\,\Omega,\quad {\mbox U}_{0}(k)+\widetilde {\mbox U}_{0}(k)=0\quad
\mbox{on}\,\,\Gamma_0
\end{equation} and solutions  $({\mbox V}_0(k),\tilde {\mbox V}_0(k))$
\begin{equation}\label{-155!!}
(2\partial_{\overline z}{\mbox V}_{0}(k) +A_2
{\mbox  V}_{0}(k), 2\partial_{ z}\widetilde {\mbox V}_{0}(k) +B_2 \widetilde
{\mbox V}_{0}(k))=0\quad\mbox{in}\,\,\Omega,\quad {\mbox V}_{0}(k)+\widetilde {\mbox V}_{0}(k)=0\quad
\mbox{on}\,\,\Gamma_0
\end{equation} for  $k\in\{1,\dots, N\}$ such that
\begin{equation}\label{1kino}
\Vert\mbox{U}_0(k)-\vec e_k\Vert_{C^{5
+\alpha}(\bar\Gamma_0)}+\Vert \tilde {\mbox V}_0(k)-\vec e_k\Vert_{C^{5
+\alpha}(\bar\Gamma_0)}\le \epsilon\quad\forall k\in\{1,\dots,N\}.
\end{equation}
This inequality and the boundary conditions in (\ref{-155!}) and in (\ref{-155!!}) imply
\begin{equation}\label{1kinogovno}
\Vert \tilde {\mbox U}_0(k)-\vec e_k\Vert_{C^{5
+\alpha}(\bar\Gamma_0)}+\Vert {\mbox V}_0(k)-\vec e_k\Vert_{C^{5
+\alpha}(\bar\Gamma_0)}\le \epsilon\quad\forall k\in\{1,\dots,N\}.
\end{equation}We define matrices $\mathcal M_1, \mathcal M_2,\mathcal R_1,\mathcal R_2$  as
\begin{eqnarray}\label{1kinodermo}
\mathcal M_1=(\tilde{\mbox U}_0(1),\dots,\tilde{\mbox U}_0(N)),\,\, \mathcal R_1=({\mbox  U}_0(1),\dots, {\mbox U}_0(N)),\nonumber\\\mathcal M_2=({\mbox V}_0(1), \dots, {\mbox V}_0(N)),\,\, \mathcal R_2=(\tilde{\mbox V}_0(1),\,\dots,\tilde {\mbox V}_0(N)).
\end{eqnarray}
By Proposition  2.3 there exists a holomorphic matrix $\mathcal Y$ such that the matrix  function $\mbox{\bf G}= \mathcal M_1\mathcal Y^{-1}\mathcal M_2^*$ solves the
partial differential equation
\begin{equation}\label{bogo}2\partial_{\bar z}\mbox{\bf G}+\mbox{\bf G}A_2^*-A_1^*\mbox{\bf G}=0\quad \mbox{in}\,\,\Omega\setminus \{x\in \bar\Omega\vert det\, \mathcal Y=0\},\quad \mbox{\bf G}\vert_{\tilde \Gamma}=I, \partial_{\vec \nu}\,\mbox{\bf G}\vert_{\tilde \Gamma}=0.
\end{equation}

Observe that the matrix ${\mbox{\bf Q}^*}^{-1} $ solves the following partial differential equation
\begin{equation}2\partial_{\bar z}{\mbox{\bf Q}^*}^{-1}+{\mbox{\bf Q}^*}^{-1}A_2^*-A_1^*{\mbox{\bf Q}^*}^{-1}=0\quad \mbox{in}\,\,\Omega \setminus \{x\in \bar\Omega\vert
det\, \mathcal P_1(x)det\,\mathcal P_2 (x)=0\},
\end{equation}
\begin{equation} {\mbox{\bf Q}^*}^{-1}\vert_{\tilde \Gamma}=I,\quad \partial_{\vec \nu}\,{\mbox{\bf Q}^*}^{-1}\vert_{\tilde \Gamma}=0.
\end{equation}

Let  matrices $\widehat{ \mathcal P_j}$ be constructed as $\mathcal P_j$ but with the different choice  of the pairs \newline $(U_0(k),\tilde U_0(k)), (V_0(k),\tilde V_0(k))$ which are  solutions to problem  (\ref{-55!}) and problem (\ref{ll1}) respectively and satisfy (\ref{kino}), (\ref{kinogovno}). In such a way  we obtain another matrix $\mbox{\bf Q}$ which satisfies to (\ref{PLP}) with possibly different set $\mathcal X.$ We denote such a matrix $\mbox{\bf Q}$ as  $\widehat{\mbox{\bf Q}}.$ By uniqueness of the Cauchy problem  for the $\partial_z$ operator
\begin{equation}\label{iii}
 \mbox{\bf Q}=\widehat { \mbox{\bf Q}}\quad \mbox{on}\,\, \Omega\setminus  \{x\in \bar\Omega\vert det\, (\mathcal P_1\mathcal P_2  \hat {\mathcal P}_1 \hat {\mathcal P}_2)(x) =0\}.
 \end{equation}
 Let $x_*\in \bar \Omega$ be a point such that $det\, (\mathcal P_1\mathcal P_2 )(x_*)=0.$ We choose the matrices $\hat {\mathcal P}_j$ such that the  determinants of these matrices are not equal to zero in some neighborhood of the point $x_*.$ Then by (\ref{iii}) the matrix ${\mbox{\bf Q}^*}^{-1}$ could be extended on the neighborhood of $x_*$ as the $C^{5+\alpha}$ matrix.
 So
 \begin{equation}2\partial_{\bar z}{\mbox{\bf Q}^*}^{-1}+{\mbox{\bf Q}^*}^{-1}A_2^*-A_1^*{\mbox{\bf Q}^*}^{-1}=0\quad \mbox{in}\,\,\Omega.
\end{equation}

By (\ref{bogo}) and uniqueness of the Cauchy problem  for the $\partial_z$ operator
$$ \mbox{\bf G}={\mbox{\bf Q}^*}^{-1}\quad\mbox{in}\,\,\Omega\setminus \{x\in \bar\Omega\vert det\,\mathcal Y=0\}.$$
Repeating the above argument we obtain that the matrix $\mbox{\bf G}^{-1}$  can be defined on $\bar \Omega$ as the function from  $C^{5+\alpha}(\bar \Omega).$ Therefore the matrix $\mbox{\bf Q}$  belongs to the space $C^{5+\alpha}(\bar\Omega)$  and solves the equation (4.16) on $\Omega$.
The operator $\tilde L_1(x,D)=\mbox{\bf Q}^{-1}L_1(x,D)\mbox{\bf Q}$ has the form
$$
\tilde L_1(x,D)=\Delta+2A_2\partial_{z}+2\tilde B_1\partial_{\bar z}+\tilde Q_1,
$$
where
$$
\tilde B_1=\mbox{\bf Q}^{-1}(B_1\mbox{\bf Q}+2\partial_{\bar z}\mbox{\bf Q}),\quad \tilde Q_1=\mbox{\bf Q}^{-1}(Q_1\mbox{\bf Q}+\Delta \mbox{\bf Q}+2A_1\partial_{z}\mbox{\bf Q}+2B_1\partial_{\bar z}\mbox{\bf Q}).
$$
The Dirichlet-to-Neumann maps of the operators $L_1(x,D)$ and $\tilde L_1(x,D)$ are the same. Let $\tilde u_1$ be the complex geometric optics solution  for the differential operator $\tilde L_1(x,D)$ constructed in the same way as solution for the operator $L_1(x,D).$ (In fact we can set $\tilde u_1=\mbox{\bf Q} u_1$
where $u_1$ be the complex geometric optics solution given by (\ref{zad}) constructed for the operator $L_1(x,D).$) For elements of the complex geometric solution $\tilde u_1$ such as $U_0,\tilde U_0, U_\tau,\tilde U_\tau$ we use the same notations as  in construction of the function $u_1.$ Since the Dirichlet-to-Neumann maps for the operators $\tilde L_1(x,D)$ and $L_2(x,D)$ are equal
there exists a  function   $ u_2$ be a solution to the following boundary value
problem:
$$
{ L}_{2}(x,D) u_2=0\quad \mbox{in}\,\,\Omega,\quad
 (\tilde u_1-u_2)\vert_{\partial\Omega}=0, \quad
\partial_{\vec\nu}(\tilde u_1-u_2)=0\quad \mbox{on}\,\,\tilde \Gamma.
$$

Setting $\tilde u=\tilde u_1-u_2, \tilde B=\tilde B_1-B_2, \tilde Q=\tilde Q_1-Q_2$ we have
\begin{equation}
{L}_2(x,{D})\tilde u
+2\tilde{\mathcal B}\partial_{\overline z}\tilde u_1
+\tilde{\mathcal Q}\tilde u_1=0 \quad \mbox{in}~ \Omega    \label{mn}
\end{equation}
and
\begin{equation}\label{mn1}
\tilde u \vert_{\partial\Omega} =0, \quad \partial_{\vec \nu}\tilde u
\vert_{\widetilde \Gamma} =0.
\end{equation}

Let $v$ be a function given by  (\ref{-3}).  Taking the scalar
product of (\ref{mn}) with $ v$  in $L^2(\Omega)$ and using
(\ref{-4}) and (\ref{mn1}), we obtain
\begin{equation}\label{ippolit1} \int_{\Omega}(2\tilde{\mathcal B}\partial_{\overline z}\tilde  u_1 +\tilde{\mathcal Q}\tilde u_1,
v) dx=\int_{\Omega}(2\tilde{\mathcal B}\partial_{\overline z}  U +\tilde{\mathcal Q} U,
V) dx+o(\frac{1}{\tau})=0,
\end{equation}
where the function  $V$ given by (\ref{gnomik1})  and

\begin{equation}
U=U_{0,\tau}e^{\tau \Phi}+\widetilde U_{0,\tau} e^{\tau \overline
\Phi}-e^{\tau\Phi}\widetilde{\mathcal R}_{\tau,  \tilde B_1}(e_1(q_1+\widetilde
q_1/\tau))-e^{\tau\overline\Phi}{\mathcal R}_{\tau, A_2}(e_1(q_2+\widetilde
q_2/\tau)).
\end{equation}

We have
\begin{proposition}\label{loop1} The following equalities are true
\begin{equation}\label{gemoroi}
 T^*_{\widetilde B_1}(\widetilde{\mathcal B}^*V_0)= T^*_{\widetilde B_1}(\bar\Phi'\widetilde{\mathcal B}^*V_0)=
\bar\Phi' T_{-B_2^*}^*(\tilde{\mathcal B}\tilde U_{0})= T_{-B_2^*}^*(\bar\Phi'\tilde{\mathcal B}\tilde U_{0})=0\quad \mbox{on}\,\,\tilde \Gamma
\end{equation}
and
\begin{equation}\label{nina}
I_{\pm,\Phi}(\tilde x)=0.
\end{equation}
\end{proposition}
{\bf Proof.}
Since the matrix $\mathcal P_1$ satisfies $2\partial_{\bar z}\mathcal P_1+A_2\mathcal P_1=0$
the matrix $\mathcal P^*_2\mathcal P_1$ is holomorphic in the domain $\Omega.$ Indeed,
\begin{equation}\label{PP}
2\partial_{\bar z}(\mathcal P^*_2\mathcal P_1)=2(\partial_{\bar z}\mathcal P^*_2\mathcal P_1+\mathcal P^*_2\partial_{\bar z}\mathcal P_1)=-\mathcal P^*_2A_2\mathcal P_1+\mathcal P^*_2A_2\mathcal P_1=0.
 \end{equation}
This implies
\begin{equation}
\int_{\partial\Omega}(\nu_1+i\nu_2)\Phi'(\mathcal P_1 {\mbox{\bf a}},\mathcal P_2 {\mbox{\bf b}})d\sigma=0,
\end{equation}

 In order to obtain the last equality we used the fact that $2\partial_{\bar z}\mathcal P^*_2=A_2^*\mathcal P^*_2.$ By (\ref{ippolit1}) the conclusion of the Proposition \ref{lodka}  holds true, if the operator $L_1(x,D)$ is replaced by the operator $\tilde L_1(x,D).$

From this equality and (\ref{1C}) we obtain

\begin{equation}\label{1qC}
\int_{\partial\Omega}(\nu_1-i\nu_2)\bar\Phi'(\mathcal C_1\bar{\mbox{\bf a}},\mathcal C_2\bar{\mbox{\bf b}})d\sigma=0,
\end{equation}
By Proposition \ref{lodka}  $\mathcal C_2^*\mathcal C_1=\tilde\Theta(\bar z)$ on $\tilde\Gamma$ where the function $\tilde\Theta$ is antiholomorphic on $\Omega.$ So
$$
\int_{\tilde \Gamma}(\nu_1-i\nu_2)\bar\Phi'(\mathcal C_2^*\mathcal C_1\bar{\mbox{\bf a}},\bar{\mbox{\bf b}})d\sigma=\int_{\tilde \Gamma}(\nu_1-i\nu_2)\bar\Phi'(\tilde\Theta\bar{\mbox{\bf a}},\bar{\mbox{\bf b}})d\sigma=-\int_{\Gamma_0}(\nu_1-i\nu_2)\bar\Phi'(\tilde\Theta\bar{\mbox{\bf a}},\bar{\mbox{\bf b}})d\sigma .
$$
We write (\ref{1qC}) as
\begin{equation}\label{11qC}
\int_{\Gamma_0}(\nu_1-i\nu_2)\bar\Phi'((\mathcal C_2^*\mathcal C_1-\tilde\Theta)\bar{\mbox{\bf a}},\bar{\mbox{\bf b}})d\sigma=0.
\end{equation}
So, by corollary 7.1 of \cite{IUY2} , from (\ref{11qC}) we obtain
\begin{equation}\label{logoped}
\mathcal C_2^*\mathcal C_1=\tilde\Theta \quad\mbox{on}\,\,\partial\Omega.
\end{equation}

We observe that for construction of $U_0$ instead of the  matrix $\mathcal C_1$ we can use $\tilde {\mathcal C}_1.$ In that case the equality (\ref{logoped}) has the form:
\begin{equation}\label{logoped1}
\mathcal C_2^*\tilde{\mathcal C}_1=\tilde\Theta_* \quad\mbox{on}\,\,\partial\Omega.
\end{equation}

We define $T^*_{\tilde B_1}(\bar\Phi'\tilde{\mathcal B}^*V_0)$ on $\Bbb R^2\setminus \bar\Omega$ by formula (\ref{giorgi1}). Now let $y=(y_1,y_2)\in \tilde \Gamma$ be an arbitrary point and $z=y_1+iy_2.$
Then, thanks to (\ref{vasilek}), for any sequence $\{y_j\}\in \Bbb R^2\setminus \bar\Omega$ such that $y_j\rightarrow  y$  we have
\begin{equation}\label{udo}
T^*_{\widetilde B_1}(\bar\Phi'\widetilde{\mathcal B}^*V_0)(y_j)\rightarrow T^*_{\widetilde B_1}(\bar\Phi'\widetilde{\mathcal B}^*V_0)(y)\quad \mbox{as}\quad j\rightarrow +\infty.
\end{equation}
Denote $z_j=y_{j,1}+iy_{j,2}.$ Indeed, by (\ref{giorgi1}) and (\ref{vasilek}) the exist a constant $C$ such that
\begin{equation}\label{chudo}
\vert T^*_{\widetilde B_1}(\bar\Phi'\widetilde{\mathcal B}^*V_0)(y_j)- T^*_{\widetilde B_1}(\bar\Phi'\widetilde{\mathcal B}^*V_0)(y)\vert\le C\int_\Omega \Vert \widetilde{\mathcal B}^*(\xi)\Vert\left\vert \frac{1}{z_j-\zeta}-\frac{1}{z-\zeta}\right\vert d\xi.
\end{equation}
Since by (\ref{vasilek}) $\Vert \widetilde{\mathcal B}^*(\xi)\Vert\vert_{\tilde\Gamma}=0$ the sequence $\left\{\Vert \widetilde{\mathcal B}^*(\xi)\Vert\left\vert \frac{1}{z_j-\zeta}-\frac{1}{z-\zeta}\right\vert\right\}$ is bounded in $L^\infty(\Omega).$
Moreover for any positive $\delta$ the above sequence converges to zero in $L^\infty(\Omega\setminus B(y,\delta)).$ Thus, from these facts and (\ref{chudo}) we have (\ref{udo}) immediately.

 By (\ref{logoped}) and (\ref{logoped1}) we have
\begin{eqnarray}\label{udo1}
T^*_{\widetilde B_1}(\bar\Phi'\widetilde{\mathcal B}^*V_0)(y_j)=\frac 12 (\mathcal C^{-1}_1 r_{0,1})(y_j)\partial^{-1}_{z} (\mathcal C_1^*\bar\Phi'\widetilde{\mathcal B}^*V_0)(y_j)
\\+ \frac 12 \widetilde{ \mathcal C}^{-1}_1(1-r_{0,1})(y_j)\partial^{-1}_{ z}
(\widetilde {\mathcal C_1}^*\bar\Phi'\widetilde{\mathcal B}^*V_0)(y_j)=\nonumber\\- \frac {1}{\pi} r_{0,1}(\overline z_j) (\mathcal C_{1}^{-1})^*(y_j)\int_{\Omega}
\frac{\bar\Phi'\partial_z(\bar\Phi'\mathcal C_1^*\mathcal C_2) \bar{\mbox{\bf  b}}}{\bar z_j-\bar \zeta}d\xi\nonumber\\-(1-r_{0,1}(\overline z_j)) (\widetilde {\mathcal C}_1^{-1})^*(y_j)
\frac {1}{\pi}\int_{\Omega} \frac{\partial_z(\bar\Phi'\tilde{\mathcal C}_1^*\mathcal C_2)\bar {\mbox{\bf b}}}{\bar z_j-\bar \zeta}d\xi=\nonumber\\- \frac {1}{4\pi} r_{0,1}(\overline z_j) (\mathcal C_{1}^{-1})^*(y_j)\int_{\partial\Omega}
\frac{(\nu_1-i\nu_2)\tilde \Theta^*\bar\Phi' \bar {\mbox{\bf b}}}{\bar z_j-\bar \zeta}d\sigma\nonumber\\-(1-r_{0,1}(\overline z_j)) (\widetilde {\mathcal C}_1^{-1})^*(y_j)
\frac {1}{4\pi}\int_{\partial\Omega} \frac{(\nu_1-i\nu_2)\widetilde \Theta^*_*\bar\Phi'\bar {\mbox{\bf b}}}{\bar z_j-\bar \zeta}d\sigma=0.\nonumber
\end{eqnarray}
Here, in order to obtain the last equality we used the fact that $z_j\notin \Omega$ and therefore the functions $\frac{1}{\bar z_j-\bar \zeta}$ are antiholomorphic on $\Omega$.
From (\ref{udo}) and  (\ref{udo1}) $T^*_{\widetilde B_1}(\bar\Phi'\widetilde{\mathcal B}^*V_0)\vert_{\tilde\Gamma}=0.$
The proof of  remaining  equalities in (\ref{gemoroi}) is the same.
Next we show that $I_{\pm,\Phi}(\tilde x)=0.$
By (\ref{begemot1}), (\ref{begemot2}) we have
\begin{eqnarray}\label{IIlin2}
I_{\pm,\Phi}(\tilde x)=\int_{\partial\Omega}\left\{ (\nu_1-i\nu_2)((2\mathcal C_2^*\mathcal C_1\mbox{\bf b}_{\pm,\tilde x}\bar\Phi', \overline{\tilde {\mbox{\bf b}}})+
(2\bar\Phi'\mathcal C_2^*\mathcal C_1\overline{\mbox{\bf a}},\widetilde{\mbox {\bf a}}_{\pm,\tilde x}))\right.\nonumber\\\left.+(\nu_1+i\nu_2)
((2 \mathcal P_2^*\mathcal P_1\mbox{\bf a}_{\pm,\tilde x}\Phi', \tilde{\mbox{\bf b}})
+(2\Phi' \mathcal P_2^*\mathcal P_1\mbox{\bf a}, \tilde{\mbox{\bf b}}_{+,\tilde x}))\right\}d\sigma.
\end{eqnarray}
Since by (\ref{logoped}) the restriction of the  function $\mathcal C_2^*\mathcal C_1$ on $\partial\Omega$  coincides with  the restriction of some antiholomorphic  in $\bar\Omega$ function and by (\ref{PP}) the restriction of the function $\mathcal P_2^*\mathcal P_1$ on $\partial\Omega$  coincides with  the restriction of some holomorphic  in $\bar\Omega$ the equality (\ref{IIlin2}) implies (\ref{nina}). The proof of thee proposition is complete. $\blacksquare$

We use the above proposition to prove the following:

\begin{proposition}\label{loop}
The following is true:
\begin{equation}\label{nok}
\bar\Phi' T^*_{\widetilde B_1}(\widetilde{\mathcal B}^*V_0)= T^*_{\widetilde B_1}(\bar\Phi'\widetilde{\mathcal B}^*V_0),
\end{equation}
\begin{equation}\label{nok1}
\bar\Phi' T_{-B_2^*}^*(\tilde{\mathcal B}\tilde U_{0})= T_{-B_2^*}^*(\bar\Phi'\tilde{\mathcal B}\tilde U_{0}).
\end{equation}
\end{proposition}

{\bf Proof.} Denote $r=\bar\Phi' T^*_{\widetilde B_1}(\widetilde {\mathcal B}^*V_0)- T^*_{\tilde B_1}(\bar\Phi'\widetilde{\mathcal B}^*V_0)$.
Then this function satisfies
$$
2\partial_{\bar z}r-\widetilde B_1^*r=0\quad \mbox{in}\,\,\Omega.
$$
By Proposition \ref{loop1}
$$
r\vert_{\tilde \Gamma}=0.
$$
By uniqueness of the Cauchy problem for the $\partial_{\bar z}$ operator $r\equiv 0.$ Proof of (\ref{nok1}) is the same.
$\blacksquare$

We use the Proposition \ref{loop} to prove the following:

\begin{proposition}\label{zanuda1}
Under conditions of Proposition \ref{lodka}  we have
\begin{eqnarray}\label{victorykk}
 -(\tilde{\mathcal B} A_2 U_0, V_0)-(\tilde Q_1(1) U_0, T^*_{\tilde B_1}(\tilde{\mathcal B}^*V_0))+(\tilde{\mathcal Q}  U_0,V_0)=0\quad\mbox{on}\,\,\Omega,
\end{eqnarray}
and
\begin{eqnarray}\label{victorykkk}(2\partial_{\bar z}\tilde{\mathcal B}\tilde U_0,\tilde  V_0)+(\tilde{\mathcal B}\tilde U_0, B_2^*\tilde V_0)- (\tilde{\mathcal Q}\tilde U_0, \tilde V_0)-(Q_1(2) \tilde V_0,T_{-B_2^*}^*(\tilde{\mathcal B}\tilde U_{0}))=0\quad\mbox{on}\,\,\Omega.\end{eqnarray}
\end{proposition}

{\bf Proof.}
We remind that $\Phi$ satisfies (\ref{zzz}), (\ref{mika}) and
\begin{equation}\label{gromila!}
\mbox{Im}\,\Phi(\widetilde x)\notin \{\mbox{Im}\,
\Phi(x); \thinspace x\in \mathcal H\setminus
\{\widetilde{x}\}\}.
\end{equation}

By Proposition \ref{lodka} equality (\ref{zaika}) holds true. Thanks to (\ref{gromila!}), (\ref{vasilek}) and Proposition \ref{loop} we can write it as
\begin{equation}\label{gromiko}
(J_\pm+ K_\pm)(\tilde x)+I_{\pm,\Phi}(\tilde x)=0.\nonumber
\end{equation}
This equality and  Proposition \ref{loop1} imply
\begin{equation}\label{gromiko}
(J_\pm+ K_\pm)(\tilde x)=0.
\end{equation}
 By Propositions \ref{osel} and \ref{loop}, we obtain

\begin{eqnarray}\label{lenta12}
\frak F_{\tau,\tilde x}(q_1 , T^*_{\tilde B_1}(\tilde B_1^*\widetilde{\mathcal A}^* V_0)-\widetilde{\mathcal A}^*V_0+2T_{\tilde B_1}^*(\partial_z \widetilde{\mathcal B}^* V_0)+T_{\tilde B_1}^*(\widetilde{\mathcal B}^*(A^*_2V_0-2\tau\bar\Phi'V_0)))=\nonumber\\
-2\tau\frak F_{\tau,\tilde x}(q_1 ,T_{\tilde B_1}^*(\widetilde{\mathcal B}^*\bar\Phi'V_0))+o(\frac{1}{\tau})=-2\tau\frak F_{\tau,\tilde x}(q_1 ,\bar\Phi'T_{\tilde B_1}^*(\widetilde{\mathcal B}^*V_0))+o(\frac{1}{\tau})=\nonumber\\
-\frac{\pi}{2\vert \mbox{det}\,\psi''(\tilde x)\vert^\frac 12}(2\partial_{\bar z} q_1,T_{\tilde B_1}^*(\widetilde{\mathcal B}^*V_0))(\tilde x)+o(\frac{1}{\tau})=\nonumber\\-\frac{\pi}{2\vert \mbox{det}\,\psi''(\tilde x)\vert^\frac 12}(\tilde Q_1(1)U_0,T_{\tilde B_1}^*(\widetilde{\mathcal B}^*V_0))(\tilde x)+o(\frac 1\tau)
\end{eqnarray}

and

\begin{eqnarray}\label{lenta22}
-2\frak F_{\tau,\tilde x}(P_{-A_2^*}^*(\widetilde{\mathcal A }(\partial_zU_0+\tau\Phi' U_0))+\widetilde{\mathcal B} \partial_{\bar z} U_{0,\tau}, q_4)=\nonumber\\-2\frak F_{\tau}(P_{-A_2^*}^*(\widetilde{\mathcal A}\tau\Phi' U_0)), q_4)+o(\frac 1\tau)=o(\frac 1\tau).
\end{eqnarray}
By (\ref{lenta12}) and (\ref{lenta22})
\begin{equation}\label{lenta3}
K_+(\tilde x)=-\frac{\pi}{2\vert \mbox{det}\,\psi''(\tilde x)\vert^\frac 12}(\tilde Q_1(1)U_0,T_{\tilde B_1}^*(\widetilde{\mathcal B}^*V_0))(\tilde x)+o(\frac 1\tau).
\end{equation}
In the similar way we compute $K_-(\tilde x):$
\begin{eqnarray}\label{LK}
\frak F_{-\tau,\tilde x} ( q_2,P_{A_2}^*(2\partial_z(\widetilde{\mathcal A}^*\tilde V_0) -\tau\Phi'2\widetilde{\mathcal A}^*\tilde V_0)-\widetilde{\mathcal B}^*\tilde V_0+P^*_{A_2}(A_2^*\widetilde{\mathcal B}^*\tilde V_0))=\nonumber\\
-2\tau\frak F_{-\tau,\tilde x} (q_2,P_{A_2}^*(\Phi'\widetilde{\mathcal A}^*\tilde V_0))+o(\frac{1}{\tau})=o(\frac{1}{\tau})
\end{eqnarray}
and
\begin{eqnarray}\label{LK1}
-2\frak F_{-\tau,\tilde x}( q_3,T_{-B_2^*}^*(2\widetilde{\mathcal A}\partial_z \tilde U_{0}+2\widetilde{\mathcal B}(\partial_{\bar z}\tilde U_{0}+\tau\bar\Phi'\tilde U_{0})))=\\
-2\frak F_{-\tau,\tilde x}( q_3,T_{-B_2^*}^*(\tau\widetilde{\mathcal B}\bar\Phi'\tilde U_{0}))+o(\frac{1}{\tau})=\nonumber
\frac{\pi}{2\vert \mbox{det}\,\psi''(\tilde x)\vert^\frac 12}(Q_1(2)\tilde V_0,T_{-B_2^*}^*(\widetilde{\mathcal B}\tilde U_{0}))+o(\frac{1}{\tau}).
\end{eqnarray}
By (\ref{LK}) and (\ref{LK1})
\begin{equation}\label{tsunamo}
K_-(\tilde x)=\frac{\pi}{2\vert \mbox{det}\,\psi''(\tilde x)\vert^\frac 12}(Q_1(2)\tilde V_0,T_{-B_2^*}^*(\widetilde{\mathcal B}\tilde U_{0}))+o(\frac{1}{\tau}).
\end{equation}
Substituting into (\ref{gromiko}) the right hand side of  formulae (\ref{lenta3}) and (\ref{tsunamo}) we obtain (\ref{victorykk}) and (\ref{victorykkk}).

Since by (\ref{bobik2})  for any $x$ from $\Omega$  exists a sequence of $x_\epsilon$  converging to $x$ we rewrite   equations (\ref{victorykk}) and (\ref{victorykkk}) as
\begin{eqnarray}\label{victory5}
 -(\tilde{\mathcal B} A_1 U_0, V_0)-(\tilde Q_1(1)U_0, T^*_{\tilde B_1}(\tilde{\mathcal B}^*V_0))+(\tilde{\mathcal Q} U_0,V_0)=0\quad\mbox{in}\,\,\,\Omega
\end{eqnarray} and
\begin{eqnarray}\label{victory911}-(2\partial_{\bar z}\tilde{\mathcal B}\tilde U_0, \tilde V_0)-(\tilde{\mathcal B}\tilde U_0, B_2^*\tilde V_0)+ (\tilde{\mathcal Q}\tilde U_0,\tilde V_0) + (Q_1(2)\tilde V_0,T_{-B_2^*}^*(\tilde{\mathcal B}\tilde U_{0}))
=0\quad\mbox{in}\,\,\,\Omega.\nonumber
\end{eqnarray}

The proof of the proposition is complete. $\blacksquare$
\bigskip
\section{\bf Step 3: End of the proof.}
\bigskip

Let $\tilde\gamma$ be a curve, without self-intersections  which pass through the point $\hat x$ and couple points $x_1, x_2$ from $\tilde\Gamma$  in such a way that the set  $\tilde \gamma\cap\partial\Omega\setminus\{x_1,x_2\}$ is empty. Denote by $\Omega_1$ a domain bounded by $\tilde\gamma$ and   part of
$\partial\Omega$ located between points $x_1$ and $x_2.$ Then we set
$\Omega_{1,\epsilon}=\{x\in\Omega\vert \thinspace dist(\Omega_1,x)<\epsilon\}.$
By Proposition \ref{nikita}  for each point $\hat x$ from $\Omega_{1,\epsilon}$ one can construct  functions $\tilde U^{(k)}_0,\tilde V^{(\ell)}_0$ satisfying  (\ref{-55!}), (\ref{ll1}) such that
$$
\tilde U^{(k)}_0(\hat x)=\vec e_k,\quad\tilde V^{(\ell)}_0(\hat x)=\vec e_\ell\quad \forall k,\ell\in \{1,\dots, N\}.
$$
Then for each $\hat x$ there exists positive $\delta(\hat x)$ such that the matrices $\{\tilde U^{(j)}_{0,i}\}$ and $\{\tilde V^{(j)}_{0,i}\}$ are invertible for any $x\in \overline{B(\hat x,\delta(\hat x))}. $ From the covering of $\bar\Omega_{1,\epsilon}$ by such a balls we take the finite subcovering $\bar \Omega_{1,\epsilon}\subset \cup_{k=1}^{\tilde N} B(x_k,\delta(x_k)).$
Then from (\ref{victory5}) we have the
differential inequality
\begin{equation}\label{london}
\vert \partial_{\bar z}\tilde{\mathcal B}_{ij}\vert \le  C_\epsilon(\sum_{k=1}^N\vert T^*_{- B_2^*}(\tilde{\mathcal B}^*\tilde U_0^{(k)})\vert+\vert\tilde{\mathcal B}\vert +\vert\tilde{\mathcal Q}\vert)\quad\mbox{in}\,\,\Omega_{1,\epsilon}, \quad \forall i,j\in \{1,\dots, N\}.
\end{equation}

Let $\phi_0\in C^2(\bar\Omega)$ be a function such that
\begin{equation}\label{indigo}
\nabla \phi_0(x)\ne 0\quad\mbox{in}\,\,\Omega_1,\quad \partial_{\tilde\nu}\phi_0\vert_{\tilde \gamma}\le \alpha' <0,\quad \phi_0\vert_{\tilde \gamma}=0,
\end{equation}
where $\tilde \nu$ is the outward normal vector to $\Omega_{1,\epsilon}$
and $\chi_\epsilon$ be a function such that
$$
 \chi_\epsilon\in C^2(\overline{\Omega_{1,\epsilon}}), \quad \chi_\epsilon=1\quad \quad\mbox{in}\,\,\Omega_1,
$$ and $\chi_\epsilon \equiv 0$ in some neighborhood of  the curve $\partial\Omega_{1,\epsilon}\setminus \tilde \Gamma.$
From (\ref{london}), (\ref{gemoroi})  we have
\begin{eqnarray}
\vert \partial_{\bar z}(\chi_\epsilon\tilde{\mathcal B}_{ij})\vert \le  C_\epsilon(\sum_{k=1}^N\vert\chi_\epsilon T^*_{-B_2^*}(\tilde{\mathcal B}^*\tilde U_0^{(k)})\vert+\vert\chi_\epsilon\tilde{\mathcal B}\vert \\+\vert[\chi_\epsilon,\partial_{\bar z}]\tilde{\mathcal B}_{ij}\vert +\vert\chi_\epsilon\tilde{\mathcal Q}\vert)\quad\mbox{in}\,\,\Omega_{1,\epsilon}, \quad \forall i,j\in \{1,\dots, N\},\nonumber\\
\chi_\epsilon\tilde{\mathcal B}\vert_{\partial\Omega_{1,\epsilon}}=\partial_{\tilde \nu}(\chi_\epsilon\tilde{\mathcal B})\vert_{\partial\Omega_{1,\epsilon}}=0.
\end{eqnarray}

Set $\psi_0=e^{\lambda\phi_0}$ with positive $\lambda$ sufficiently large.
Applying the Carleman estimate to the above inequality we have
\begin{eqnarray}\label{volk1}
\int_{\Omega_{1,\epsilon}} e^{2\tau\psi_0}(\frac 1\tau\vert \nabla \chi_\epsilon \tilde{\mathcal B}\vert^2+\tau\vert \chi_\epsilon\tilde{\mathcal B}\vert^2) dx\le C \int_{\Omega_{1,\epsilon}} (\sum_{k=1}^N\vert \chi_\epsilon  T^*_{-B_2^*}(\tilde{\mathcal B}^*\tilde U_0^{(k)})\vert^2 \nonumber\\ +\chi_\epsilon^2(
\vert \tilde{ \mathcal B}\vert^2+\vert\tilde{\mathcal Q}\vert^2)+\vert [\chi_\epsilon,\partial_{\bar z}]  \tilde{\mathcal B}\vert^2 ) e^{2\tau\psi_0}dx\quad\forall\tau\ge \tau_0.
\end{eqnarray}

By the Carleman estimate for the operator $\partial_z$ and (\ref{gemoroi})  there exist $C$ and $\tau_0$ independent of $\tau $ such that
\begin{equation}\label{volk2}\int_{\Omega_{1,\epsilon}}\vert \chi_\epsilon  T^*_{-B^*_2}(\tilde{\mathcal B}^*\tilde U_0^{(k)})\vert^2 e^{2\tau\psi_0}dx\le C\int_{\Omega_{1,\epsilon}}(\vert[\chi_\epsilon ,\partial_z] T^*_{-B^*_2}(\tilde{\mathcal B}^*\tilde U_0^{(k)})\vert^2+\vert\chi_\epsilon\tilde{\mathcal B}^*\tilde U_0^{(k)}\vert^2) e^{2\tau\psi_0}dx
\end{equation}
and
\begin{equation}\label{volk3}\int_{\Omega_{1,\epsilon}}\vert \chi_\epsilon   T^*_{\tilde B_1}(\tilde{\mathcal B}^* V_0^{(k)})\vert^2 e^{2\tau\psi_0}dx\le C\int_{\Omega_{1,\epsilon}}(\vert[\chi_\epsilon ,\partial_{ z}]  T^*_{\tilde B_1}(\tilde{\mathcal B}^* V_0^{(k)})\vert^2+\vert \chi_\epsilon\tilde{\mathcal B}^*V_0^{(k)}\vert^2) e^{2\tau\psi_0}dx
\end{equation}
for all $\tau\ge\tau_0.$

Combining estimates  (\ref{volk1}),  (\ref{volk2}) we obtain that there exist a constant $C$ independent of $\tau$ such that
\begin{eqnarray}\label{volk4!}
\int_{\Omega_{1,\epsilon}} e^{2\tau\psi_0}(\frac 1\tau\vert \nabla (\chi_\epsilon\tilde{ \mathcal B})\vert^2+\tau\vert \chi_\epsilon\tilde{\mathcal B}\vert^2) dx\\
 \le C \int_{\Omega_{1,\epsilon}}(\nonumber
  \chi_\epsilon^2(
\vert \tilde{ \mathcal B}\vert^2+\vert\tilde{\mathcal Q}\vert^2)+\sum_{k=1}^N\vert[\chi_\epsilon ,\partial_z] T^*_{-B^*_2}(\tilde{\mathcal B}^*\tilde U_0^{(k)})\vert^2+
 \vert [\chi_\epsilon,\partial_{\bar z}]\tilde{\mathcal B}\vert^2 ) e^{2\tau\psi_0}dx \quad \forall\tau\ge\tau_0.\nonumber
\end{eqnarray}

 For all sufficiently large $\tau$  the term $\int_{\Omega_{1,\epsilon}}\vert \chi_\epsilon\widetilde{\mathcal B}\vert^2e^{2\tau\psi_0}dx$ absorbed by the integral on the left hand side. Moreover, thanks to the choice of the  function $\chi_\epsilon,$ we have supports of coefficients for the operator  $ [\chi_\epsilon ,\partial_{\bar z}]$  are located in the domain $\Omega_{1,\epsilon}\setminus\Omega_{1,\frac{\epsilon}{2}}.$
Hence one can write the estimate (\ref{volk4!}) as

\begin{eqnarray}\label{volk4}
\int_{\Omega_{1,\epsilon}} e^{2\tau\psi_0}(\frac 1\tau\vert  \nabla (\chi_\epsilon\tilde{\mathcal B})\vert^2+\tau\vert \chi_\epsilon\tilde{\mathcal B}\vert^2) dx\le C\int_{\Omega_{1,\epsilon}}\chi^2_\epsilon\vert \tilde{\mathcal Q}\vert^2e^{2\tau\psi_0}dx\\
+ C\int_{\Omega_{1,\epsilon}\setminus \Omega_{1,\frac{\epsilon}{2}}}
 (\sum_{k=1}^N \vert[\chi_\epsilon ,\partial_z] T^*_{-B^*_2}(\tilde{\mathcal B}^*\tilde U_0^{(k)})\vert^2+
\vert [\chi_\epsilon,\partial_{\bar z}]\tilde{\mathcal B}\vert^2 ) e^{2\tau\psi_0}dx\quad \forall\tau\ge\tau_1.\nonumber
 \end{eqnarray}

By Proposition \ref{nikita}  for each point $\hat x$ from $\Omega$ one can construct such a function $U^{(k)}_0, V^{(\ell)}_0$ satisfying  (\ref{-55!}), (\ref{ll1}) such that
$$
 U^{(k)}_0(\hat x)=\vec e_k,\quad V^{(\ell)}_0(\hat x)=\vec e_\ell\quad \forall k,\ell\in \{1,\dots,\N\}.
$$
Then for each $\hat x\in\bar\Omega_{1,\epsilon}$ there exists positive $\delta(\hat x)$ such that the matrices $\{ U^{(j)}_{0,i}\}$ and $\{ V^{(j)}_{0,i}\}$ are invertible for any $x\in \overline{B(\hat x,\delta(\hat x))}. $ From the covering of $\Omega_{1,\epsilon}$ by such a balls we take the finite subcovering $\bar \Omega\subset \cup_{k=\tilde N}^{\tilde N+N^*} B(x_k,\delta(x_k)).$ Then there exists $C_\epsilon$ such that
\begin{equation}\label{volk44}
\vert\tilde{ \mathcal Q}\vert\le  C_\epsilon(\vert\tilde{\mathcal B}\vert+\sum_{k=\tilde N+1}^{\tilde N+N^*}\vert T^*_{\tilde B_1}(\tilde{\mathcal B}^*V_0^{(k)})\vert)\quad\mbox{in}\,\,\Omega_{1,\epsilon}.
\end{equation}

Combining   (\ref{volk3}), (\ref{volk4}) and (\ref{volk44}) we obtain that there exists a constant $C_5$ independent of $\tau$
\begin{eqnarray}\label{volk41}
\int_{\Omega_{1,\epsilon}} e^{2\tau\psi_0}(\frac 1\tau\vert \nabla (\chi_\epsilon\tilde{ \mathcal B})\vert^2+\tau\vert \chi_\epsilon\tilde{\mathcal B}\vert^2) dx\le C_5\int_{\Omega_{1,\epsilon}\setminus \Omega_{1,\frac{\epsilon}{2}}}
(\sum_{k=1}^N\vert[\chi_\epsilon ,\partial_z] T^*_{-B^*_2}(\mathcal B^*\tilde U_0^{(k)})\vert^2\nonumber\\
 +\sum_{k=\tilde N+1}^{\tilde N+N^*}\vert [\chi_\epsilon ,\partial_z] T^*_{\tilde B_1}(\tilde{\mathcal B}^*V_0^{(k)})\vert+
\vert [\chi_\epsilon,\partial_{\bar z}]\tilde{\mathcal B}\vert^2 ) e^{2\tau\psi_0}dx\quad \forall\tau\ge\tau_1.
 \end{eqnarray}
 By (\ref{indigo}) for all sufficiently small positive $\epsilon$ there exists a positive constant $\alpha<1$ such that
 \begin{equation}\label{soroka}
 \psi_0(x) <\alpha\quad\mbox{on} \quad\Omega_{1,\epsilon}\setminus\Omega_{1,\frac{\epsilon}{2}}.
 \end{equation}
  Since $\hat x\in \mbox{supp}\, \tilde{\mathcal B}\cap \tilde\gamma$  and  thanks to the fact $\partial_{\tilde\nu}\phi_0\vert_{\tilde \gamma}\le \alpha ' <0$ there exists $\kappa>0$ such that
 \begin{equation}\label{volk6}\kappa e^\tau\le \int_{\Omega_{1,\epsilon}} e^{2\tau\psi_0} \vert \chi_\epsilon\tilde{\mathcal B}\vert^2e^{2\tau\psi_0} dx \quad \forall\tau\ge\tau_1.
 \end{equation}
  By (\ref{soroka}) we can estimate the right hand side of the inequality (\ref{volk4}) as
  \begin{eqnarray}\label{volk5}
 C_5\int_{\Omega_{1,\epsilon}\setminus \Omega_{1,\frac{\epsilon}{2}}}
(\sum_{k=1}^N\vert[\chi_\epsilon ,\partial_z] T^*_{-B^*_2}(\tilde{\mathcal B}^*\tilde U_0^{(k)})\vert^2
 +\sum_{k=\tilde N+1}^{\tilde N+N^*}\vert [\chi_\epsilon ,\partial_z]T^*_{\tilde B_1}(\tilde{\mathcal B}^*V_0^{(k)})\vert\nonumber\\+
\vert [\chi_\epsilon,\partial_{\bar z}]\tilde{\mathcal B}\vert^2 ) e^{2\tau\psi_0}dx\le C_6 e^{\alpha\tau}\quad \forall\tau\ge\tau_1,
\end{eqnarray} where $C_5,C_6$ are positive constants independent of $\tau.$
Using (\ref{volk6}) and (\ref{volk5}) in (\ref{volk4}) we obtain
$$
\kappa e^\tau \le C_7 e^{\alpha\tau}\quad \forall\tau\ge\tau_1.
$$
Since $\alpha<1$ we arrived to the contradiction. Hence
$$
\tilde{\mathcal B}=\tilde{\mathcal Q}=0\quad \mbox{on}\,\, \Omega\setminus \mathcal X_{\epsilon_0}.
$$
The proof of theorem is complete. $\blacksquare$

\end{document}